# Machine Learning in Electronic Quantum Matter Imaging Experiments

Yi Zhang[1,*], A. Mesaros[1,2,*], K. Fujita[3], S.D. Edkins[1,4], M.H. Hamidian[1,5], K. Ch'ng[6], H. Eisaki[7], S. Uchida[7,8], J.C. Séamus Davis[1,3,9,10], E. Khatami[6] and Eun-Ah Kim[1]

[1] *Department of Physics, Cornell University, Ithaca, NY 14853, USA.*
[2] *Laboratoire de Physique des Solides, Université Paris-Sud, CNRS, 91405 Orsay Cedex, France.*
[3] *CMPMS Department, Brookhaven National Laboratory, Upton, NY 11973, USA.*
[4] *Department of Applied Physics, Stanford University, Stanford, CA 94305, USA.*
[5] *Department of Physics, Harvard University, Cambridge, MA 02138, USA.*
[6] *Department of Physics and Astronomy, San Jose State University, San Jose, CA 95192, USA.*
[7] *AIST, Tsukuba, Ibaraki 305-8568, Japan.*
[8] *Department of Physics, University of Tokyo, Bunkyo-ku, Tokyo 113-0033, Japan.*
[9] *Department of Physics, University College Cork, Cork T12R5C, Ireland*
[10] *Clarendon Laboratory, Oxford University, Oxford, OX1 3PU, UK*
* These authors contributed equally to this work.

**Essentials of the scientific discovery process have remained largely unchanged for centuries[1]: systematic human observation of natural phenomena is used to form hypotheses that, when validated through experimentation, are generalized into established scientific theory. Today, however, we face major challenges because automated instrumentation and large-scale data acquisition are generating data sets of such volume and complexity as to defy human analysis. Radically different scientific approaches are needed, with machine learning (ML) showing great promise, not least for materials science research[2-5]. Hence, given recent advances in ML analysis of synthetic data representing electronic quantum matter (EQM)[6-16], the next challenge is for ML to engage equivalently with experimental data. For example, atomic-scale visualization of EQM yields arrays of complex electronic structure images[17], that frequently elude effective analyses. Here we report development and training of an array of artificial neural networks (ANN)**

**designed to recognize different types of hypothesized order hidden in EQM image-arrays. These ANNs are used to analyze an experimentally-derived EQM image archive from carrier-doped cuprate Mott insulators. Throughout these noisy and complex data, the ANNs discover the existence of a lattice-commensurate, four-unit-cell periodic, translational-symmetry-breaking EQM state. Further, the ANNs find these phenomena to be unidirectional, revealing a coincident nematic EQM state. Strong-coupling theories of electronic liquid crystals[18,19] are congruent with all these observations.**

*1* Frontier research in EQM concentrates on exotic electronic phases that emerge when electrons interact so strongly that they lack a definite momentum. These electrons often self-organize into complex new states of EQM including, for example, electronic liquid crystals[18,19], high temperature superconductors[20,21], fractionalized electronic fluids and quantum spin liquids. In this field, vast experimental data sets have emerged, for example from real space (***r***-space) visualization of EQM using spectroscopic imaging scanning tunneling microscopy[17] (SISTM), from momentum space (***k***-space) visualization of EQM using angle resolved photoemission (ARPES), or from modern X-ray[22] and neutron scattering. The challenge is to develop ML strategies capable of scientific discovery using such large and complex experimental data structures from EQM experiments.

*2* An excellent example is the electronic structure of the $CuO_2$ plane in the cuprate compounds supporting high temperature superconductivity[20] (Fig. 1a). With one electron per Cu site, strong Coulomb interactions produce charge localization in an antiferromagnetic Mott insulator (MI) state. Removing $p$ electrons (adding $p$ 'holes') per $CuO_2$ plaquette generates the 'pseudogap' (PG) phase[20]. It exhibits strongly depleted density-of-electronic states $N(E)$ for energies $|E| < \Delta_1$, where $\Delta_1$ is the characteristic pseudogap energy scale that emerges for $T < T^*(p)$ (Fig. 1a). Although the PG phase has defied microscopic identification for decades[20], recently it has been reported that rotational and translational symmetry are spontaneously broken in this phase. Rotational symmetry breaking is referred to as a nematic (NE) state[18,19,23,24]; it occurs at wavevector $\boldsymbol{Q}$=0 as the breaking of 90°-rotational ($C_4$) symmetry at $T < T^*(p)$ (Fig. 1a). This presents a conundrum because, in theory, ordering at $\boldsymbol{Q}$=0 cannot open an energy gap in the electronic spectrum. The translational symmetry breaking or density wave (DW) state, which should open such an energy gap, is detected using SISTM visualization[17] and X-ray scattering[22]. It consists of periodic spatial modulations of electronic structure with finite wavevector $\boldsymbol{Q}$ and thus with periodicity $\lambda = 2\pi/|\boldsymbol{Q}|$, that occur within the pseudogap phase (Fig. 1a). A key challenge for this field is to identify the correct microscopic theory for the DW state (Methods Section 1), and to find the relationship (if any) between it and both the NE state and the pseudogap.

3 A DW state with wavevector $\boldsymbol{Q}$ is described by a spatially modulating function $A(\boldsymbol{r}) = D(\boldsymbol{r})Cos(\boldsymbol{Q}\cdot\boldsymbol{r} + \phi_0(\mathbf{r}))$ : $A(\boldsymbol{r})$ represents the density amplitude, $\phi_0(\boldsymbol{r})$ represents effects of disorder and topological defects, $\lambda = 2\pi/|\boldsymbol{Q}|$ is the periodicity, $\boldsymbol{Q}/|\boldsymbol{Q}|$ is the direction of the modulation, while $D(\boldsymbol{r})$ is the DW form factor symmetry. For a tetragonal crystal, an $s$-symmetry form factor remains unchanged under 90° rotations, while a $d$-symmetry form factor changes sign as observed in cuprates[25]. One theoretical approach to understanding a DW state is based on conventional electrons with well-defined wave momentum $\boldsymbol{p}(E) = \hbar\boldsymbol{k}(E)$. DW states can then appear at a wavevector $\boldsymbol{Q} = (\boldsymbol{k}_i(E=0) - \boldsymbol{k}_f(E=0))$ if many pairs of $(\boldsymbol{k}_i(0), \boldsymbol{k}_f(0))$ are connected by the same wavevector $\boldsymbol{Q}$, i.e., nested (red arrow Fig. 1b). Under these circumstances, $\boldsymbol{Q}$ should usually be incommensurate (Fig. 1b). Alternatively, strongly interacting particle-like electrons may have well-defined position in $\boldsymbol{r}$-space, being fully localized in the MI phase or self-organized into electronic liquid crystal states[18,19,24]. For cuprates, such states are often predicted[18,19,24] to exhibit periodic charge density modulations that are unidirectional, crystal-lattice-commensurate, with wavelength $\lambda = 4a_0$ or wavevector $\boldsymbol{Q} = 2\pi/a_0\,(0.25,0)$ oriented along the Cu-O-Cu axis (Fig. 1c and Methods Section 1). Such lattice-commensurate charge modulations in position-based theories (Fig. 1c) are expected to be robust against changes with electron-density $p$ and electron-energy, while those associated with the geometry of Fermi surface in momentum-based theories (Fig. 1b) are expected to evolve continuously with $p$.

***4*** A central challenge has therefore been to determine if the electronic structure modulations in hole-doped $CuO_2$ (e.g. Fig. 1d,e) are lattice-commensurate, unidirectional, with specific periodicity, or if they evolve continuously with electron-density and electron-energy. But, because of their inherent limitations, it has not been possible to discriminate between these position-based or momentum-based theoretic perspectives by using traditional analysis techniques. First, due to the extreme disorder observed in cuprate EQM images[17] (Fig. 1d) or concomitantly the broad line-widths detected in reciprocal space[22], theory demonstrates that conventional Fourier analysis is fundamentally limited[26,27] in determining the exact symmetries of the EQM state. Second, when such complicated electronic-structure motifs exist at atomic-scale in $\boldsymbol{r}$-space[17], Fourier analysis spreads all that information throughout reciprocal space. Consequently, the customary Fourier analysis of SISTM and X-ray data focusing on a single intensity peak, which has long reported incommensurate modulations that evolve continuously with $p$ in the range $0.22 \lesssim Q(2\pi/a_0) \lesssim 0.3$ (Ref. 17,22), disregards much information. Specifically, the key insights contained in atomic-scale electronic-structure motifs (Fig. 1d), discommensurations[28] and topological defects (Methods Section 2) are all discarded. By contrast, ML analysis of EQM images holds great promise because it avoids this information loss and analyzes the complete image array objectively.

***5*** High-data-volume imaging studies of EQM (e.g. Fig. 1e) use SISTM, a technique for visualizing $N(\boldsymbol{r}, E)$ with sub-atomic resolution and crystal-lattice register[17]. The resulting image-array for a given sample is built up from measurements of STM-tip-sample differential electron tunneling conductance $dI/dV(\boldsymbol{r}, V) \equiv g(\boldsymbol{r}, V)$ at a square array of locations $\boldsymbol{r}$ and at a range of tip-sample voltage differences $V$. For technical reasons, images $Z(\boldsymbol{r}, V) \equiv g(\boldsymbol{r}, +V)/g(\boldsymbol{r}, -V)$, which accurately represent the spatial symmetry of electronic structure but avoid systemic errors[17], are most frequently used. While Fourier analysis of $Z(\boldsymbol{r}, V)$ to yield $Z(\boldsymbol{q}, V)$ is an obvious approach to studying the EQM modulation wavevectors[17,22], it faces severe limitations as discussed above. To identify the fundamental broken-symmetry EQM state from an array of such $Z(\boldsymbol{r}, E = eV)$ images (e.g. Fig. 1e) therefore poses an iconic challenge for ML techniques.

*6* Here we introduce a specific ML approach using ANN's to achieve hypothesis testing with EQM image-arrays. It is based upon supervised ML within an ANN-human coalition. Its goals are to automatically search experimental EQM image-arrays (e.g. Fig. 1e), to recognize spatial modulations in a variety of distinct categories, to identify their fundamental periodicity and lattice register throughout an image, and to distinguish if the modulations are unidirectional or bidirectional. The first stage is generation of sets of ANN training images, each labeled by a hypothesis: the different DW modulations to be discerned. Here, we test four hypotheses associated with four distinct types of ideal periodic modulations, all with a d-symmetry form factor, and with fundamental wavelengths $\lambda$=4.348$a_0$, 4.000$a_0$, 3.704$a_0$, 3.448$a_0$ respectively. Notice that only category 2 represents a commensurate pattern with $\lambda = 4a_0$. Four training sets for categories $C$=1,2,3,4 are then generated using identical procedures, in which we introduce specific forms of heterogeneity designed to mimic the noise, intrinsic disorder and topological defects of experimental data (Fig. 2a and Methods Section 3). Throughout these simulated training-image-sets, the heterogeneity disrupts the long-range ordered patterns in $\boldsymbol{r}$-space, as shown for a typical training image in Fig. 2b. It also scrambles the peaks in the d-symmetry Fourier transforms[17] of the training images, rendering them broad and chaotic (Fig. 2c). In the second stage, we establish an ANN architecture that trains well with these training-image-sets. During training, the parameters of the ANN are adjusted iteratively to minimize a cross-entropy cost function[29]. Stochastic gradient descent along with backpropagation[30] is used for lowering the cost function. The training is complete and all parameters of each ANN are set when the cross-entropy[31] saturates. Each finalized ANN generally has an accuracy >99% when tested on validation images (Fig. 2d and Methods Section 4). The ANN design is a fully connected feed forward network with a single hidden layer (Fig. 3 and Methods Section 4). Statistical reliability of this ML system against different network architectures and different initial conditions is achieved by training 81 distinct ANNs in parallel with the same training image-set (Methods Section 4).

7 Our ANN ensemble is first used to hypothesis test the experimental EQM image-arrays versus changing electron-density. The measured $Z(\boldsymbol{r}, E)$ electronic-structure images are from samples of the hole-doped cuprate $Bi_2Sr_2CaCu_2O_8$ that span the range $0.06 \leq p \leq 0.20$. Obviously disorder and complexity of EQM abound in $Z(\boldsymbol{r}, \Delta_1)$ throughout this whole electron-density range (black double headed arrow in Fig. 1a) and are equally apparent in the broad fluctuating peaks around $(Q_x \pm \delta Q_x, \delta Q_y)2\pi/a_0$ and $(\delta Q_x, Q_y \pm \delta Q_y)2\pi/a_0$ in $Z(\boldsymbol{q}, \Delta_1)$ (see Figs. 3a,b). Definite fundamental periodicities seem undetectable in these $Z(\boldsymbol{r}, \Delta_1)$ data. The set of experimental $Z(\boldsymbol{r}, E)$ image-arrays have FOV 16nmX16nm, but are measured in a sequence of independent experiments on distinct crystals with $p$≈0.06, 0.08, 0.085, 0.14, 0.20 ($T_c$(K)=20, 45, 50, 74, 82). The ANNs analyze these $Z(\boldsymbol{r}, \Delta_1)$ images as a function of $p$, focusing on the pseudogap energy $E = \Delta_1(p)$ because cuprate EQM symmetry-breaking emerges at this energy[17,25]. Figures 4a-e show the actual $Z(\boldsymbol{r}, \Delta_1)$ images presented to the trained ANN system while Figs. 4f-j show their $d$-symmetry Fourier transforms. The ANN's succeed with high reliability in discriminating and identifying the periodic motifs throughout these images (Methods Section 5). In Figures 4k-

o we show the response of the ANNs as the probability $P(C)$ that the presented EQM image is identified in the category $C$. Here the ANNs reveal that, on the average, the phenomenology of the $C$=2, $\lambda = 4a_0$ training-images has the highest probability of being recognized within the $Z(\boldsymbol{r}, \Delta_1)$ image array, but only for electron-densities $0.06 \leq p \leq 0.14$. Thus, the ANNs identify a predominant translational symmetry breaking, occurring commensurately with the specific wavelength $\lambda = 4a_0$ (Fig. 4a-d). Overall, the ANNs conclude that the identical, commensurate, $4a_0$ periodic, electronic structure modulations were hidden throughout the $E \approx \Delta_1$ EQM images from the $0.06 \leq p \leq 0.14$ area of the $CuO_2$ phase diagram.

**8** A second key physics issue is the energy dependence within an $Z(\boldsymbol{r}, E)$ image-array. Quasiparticle scattering interference[17] (QPI) occurs when an impurity atom scatters wave-like states $\boldsymbol{k}_i(E)$ into $\boldsymbol{k}_f(E)$, resulting in quantum inference at wavevectors $\boldsymbol{Q}_{if}(E) = \boldsymbol{k}_i(E) - \boldsymbol{k}_f(E)$, and generating modulations of $N(\boldsymbol{r}, E)$ or its Fourier transform $N(\boldsymbol{Q}_{if}, E)$. QPI is a distinct physical phenomenon from a DW state because, while the modulation wavevectors of the former evolve rapidly with $E$, for latter they do not. Therefore, the ANNs explore a $Bi_2Sr_2CaCu_2O_8$ $Z(\boldsymbol{r}, E)$ array of 16nmX16nm EQM images, that are measured in a sequence of independent experiments at distinct electron-energy $E$=*66, 96, 126, 150*(meV) on the same crystal with $p$=0.08. Figures 5a-d show this $Z(\boldsymbol{r}, E)$ image set that is presented to the same ANN system. EQM complexity in the identical field of view now evolves rapidly with electron-energy because they are dominated by QPI. Similarly, Figures 5e-h are the $d$-symmetry Fourier transforms $Z(\boldsymbol{q}, E)$ from Figures 5a-d, showing broad fluctuating peaks that evolve rapidly with electron-energy as expected in QPI. Well-defined fundamental periodicities appear indiscernible in these $Z(\boldsymbol{r}, E)$ (A-D); $Z(\boldsymbol{q}, E)$ (E-H) data. However, Figures 5j-l demonstrate that the ANN suite finds the hypothesis category with the highest recognition probability to again be $C$=2, meaning that the predominant modulations have period $4a_0$ for all energies exceeding 66meV (Fig. 5b-d). Again, despite intense masking by QPI phenomena, the ANN's recognize commensurate, $4a_0$ periodic, DW modulations and reveal that it occurs predominantly near the pseudogap energy scale $E = \Delta_1$.

9 A third ANN discovery in Fig. 5i-l is that the commensurate, $4a_0$ periodic modulations exhibit a strong preference for breaking symmetry under $90^o$ rotations ($C_4$). This is revealed because the ANN array yields up to 3 times higher probability in the specific category ($C$=2) when the data is presented in the X orientation (red) compared to when the identical data is presented to it in the Y orientation (yellow) (Fig. 5j-l). Although the extreme nanoscale disorder masks it in the images Fig. 5a-d, the DW modulations are therefore occurring primarily along the x-axis of the $CuO_2$ plane. ANN analysis of the energy dependence of this complete $Z(\boldsymbol{r}, E)$ image array in Extended Data Fig. 1 further confirms that the appearance of this nematicity (Fig. 5i-l) occurs approaching the pseudogap energy scale which is $|\Delta_1| \approx 80meV$. Thus, the ANNs find that a nematic state emerges at the pseudogap energy specifically due to highly disordered yet unidirectional $4a_0$ periodic modulations. This discovery strongly implies that the nematic electronic structure of $CuO_2$ is a vestigial nematic state[32] whose characteristic energy gap is the pseudogap. Advanced theory predicts that a unidirectional DW that is reduced by disorder to extremely short spatial coherent lengths, should generate a nematic state dubbed a vestigial nematic state[32]. Although experimental validation for this hypothesis is formally impossible using conventional FT techniques[26,27], here it is demonstrably achievable by an ANN array (Fig. 5, Extended Data Fig. 1). Existence of a vestigial nematic state in carrier-doped $CuO_2$ would provide a direct, internally consistent link between a nematic state and the unidirectional $4a_0$ periodic DW modulations, whose energy gap is the pseudogap (Fig. 4). The evidence for a vestigial nematic emerged unexpectedly from ANN analysis of experimental image arrays not optimized for such studies; for the ANN suite to determine a complete $p$ dependence will require new measurements of appropriately optimized image arrays.

*10* To summarize: we have developed and demonstrated a new general protocol for ML-based identification of the symmetry-breaking ordered states in electronic structure image-arrays from EQM visualization experiments. Our ANNs are trained to learn the defining motifs of each category including its topological defects, and to recognize those motifs in real EQM image arrays (Fig. 1e). Despite the complexity of the hole-doped Mott insulator state, instrument distortion and noise, and the intense electronic disorder of the EQM image arrays studied (Figs.1d,3a,b;4,5), the ANNs repeatedly and reliably discover predominant features of a specific ordered state. Its signature, for $0.06 \leq p \leq 0.14$, is a lattice-commensurate, unidirectional, $d$-symmetry form factor, $\lambda = 4a_0$ periodic electronic structure modulation (Fig. 4). As an advance in CM physics, the predominance of this phenomenology (Fig. 4) implies that a strong coupling position-based theory is central to these broken-symmetry states of carrier-doped $CuO_2$. The ANN array also reveals evidence that it is the $\lambda = 4a_0$ DW modulations at the pseudogap energy that break the global rotational symmetry to generate a nematic state (Fig. 5, Extended Data Fig. 1). This implies that the PG region of the $CuO_2$ phase diagram (Fig. 1a) contains a vestigial nematic state. Concurrently, a milestone for general scientific technique is achieved with the demonstration that ANN's can process and identify specific broken symmetries of highly complex image-arrays from non-synthetic experimental EQM data. Overall, these combined advances open the immediate prospect of additional ML-driven scientific discovery in EQM studies.

## References

[1] Bacon, Francis, *The Advancement of Learning* (1605); republished Paul Dry Books (2001).

[2] Ouyang, R. et al. SISSO: a compressed-sensing method for identifying the best low-dimensional descriptor in an immensity of offered candidates, *Phys. Rev. Materials* **2**, 083802 (2018).

[3] Stanev, V. et al. Machine learning modeling of superconducting critical temperature, *Computational Materials* **4**, 29 (2018).

[4] Rosenbrock, C. W., Homer, E. R., Csányi, G. and Hart, G. L. W. Discovering the building blocks of atomic systems using machine learning: application to grain boundaries. *npj Computational Materials* **3**, 29 (2017).

[5] Kusne A.G. et al. On-the-fly machine-learning for high-throughput experiments: search for rare-earth-free permanent magnets, *Scientific Reports* **4**, 6367 (2014).

[6] Carrasquilla, J. and Melko, R. G. Machine learning phases of matter. *Nature Physics* **13**, 431(2017).

[7] Carleo, G. and Troyer, M. Solving the quantum many-body problem with artificial neural networks. *Science* **355**, 602 (2017).

[8] Torlai, G. and Melko, R.G. Neural decoder for topological codes. *Phys. Rev. Lett.* **119**, 030501 (2017).

[9] Van Nieuwenburg, E. P. L., Liu, Y.-H. and Huber, S. D. Learning phase transitions by confusion. *Nature Physics* **13**, 435-439 (2017).

[10] Broecker, P., Carrasquilla, J., Melko, R. G. and Trebst, S. Machine learning quantum phases of matter beyond the fermion sign problem. *Scientific Reports* **7**, 8823 (2017).

[11] Ch'ng, K., Carrasquilla, J., Melko, R. G. and Khatami, E. Machine Learning Phases of Strongly Correlated Fermions. *Phys. Rev. X* **7**, 031038 (2017).

[12] Zhang, Y. and Kim, E.-A. Quantum loop topography for machine learning. *Phys. Rev. Lett.* **118**, 216401 (2017).

[13] Deng, D.-L. , Li, X. and Das Sarma, S. Quantum entanglement in neural network states. *Phys. Rev. X* **7**,021021 (2017).

[14] Stoudenmire, E. M. and Schwab, D. J. Supervised learning with tensor networks. *Advances in Neural Information Processing Systems (NIPS)* **29**, 4799 (2016).

[15] Schindler, F., Regnault, N. and Neupert, T. Probing many-body localization with neural networks. *Phys. Rev. B* **95**, 245134 (2017).

[16] Torlai, G., Mazzola, G., Carrasquilla, J., Troyer, M., Melko, R. G. and Carleo, G. Neural-network quantum state tomography. *Nature Physics* **14**, 447 (2018).

[17] Fujita, K. *et al*, *Spectroscopic Imaging STM*: *Atomic-Scale Visualization of Electronic Structure and Symmetry in Underdoped Cuprates*, In *Strongly Correlated Systems - Experimental Techniques,* ed. Avella, A. and Mancini, F. pp 73-109. Springer-Verlag Berlin Heidelberg (2015).

[18] Kivelson, S. A., Fradkin, E. and Emery, V. J. Electronic liquid-crystal phases of a doped Mott insulator. *Nature* **393**, 550 (1998).

[19] Zaanen, J. Self-organized one dimensionality. *Science* **286**, 251 (1999).

[20] Keimer, B., Kivelson, S. A., Norman, M. R., Uchida, S. and Zaanen, J. From quantum matter to high-temperature superconductivity in copper oxides. *Nature* **518**, 179–186 (2015).

[21] Wang, F. and Lee, D.-H. The Electron-Pairing Mechanism of Iron-Based Superconductors. *Science* **332**, 200-204 (2011).

[22] Comin, R. and Damaschelli, A. Resonant X-Ray Scattering Studies of Charge Order in Cuprates, *Annu. Rev. Condens. Matter Phys.* **7**:369-405 (2016).

[23] Fradkin, E. *et al.* Nematic Fermi Fluids in Condensed Matter Physics. *Annual Review of Condensed Matter Physics* **1**, 153-178 (2010).

[24] Fradkin, E., Kivelson, S. A. and Tranquada, J. M. Colloquium: Theory of intertwined orders in high temperature superconductors. *Rev. Mod. Phys.* **87**, 457 (2015).

[25] Hamidian, M. H. *et al.* Atomic-scale electronic structure of the cuprate d-symmetry form factor density wave state. *Nature Physics* **12**, 150 (2016).

[26] Robertson, J. A. *et al.* Distinguishing patterns of charge order: Stripes or checkerboards. *Phys. Rev. B* **74**, 134507 (2006).

[27] Del Maestro, A., Rosenow, B. and Sachdev, S. From stripe to checkerboard ordering of charge-density waves on the square lattice in the presence of quenched disorder. *Phys. Rev.* ***B*** 74, 024520 (2006).

[28] Mesaros, A. et al. Commensurate 4a0-period charge density modulations throughout the $Bi_2Sr_2CaCu_2O_{8+x}$ pseudogap regime. *Proc. Nat'l. Acad. Sci.* **113**, 12661-12666 (2016).

[29] Nielsen, M. A. *Neural Networks and Deep Learning* (Determination Press, 2015); http://neuralnetworksanddeeplearning.com.

[30] Rumelhart, D. E., Hinton, G. E. and Williams, R. J. Learning representations by back-propagating errors. *Nature* **323**, 533-536 (1986).

[31] Cover, T. M., and Thomas, J. A. *Elements of Information Theory* (2nd Edition, Wiley, 1991).

[32] Nie, L. et al. Quenched disorder and vestigial nematicity in the pseudogap regime of the cuprates *Proc. Nat'l. Acad. Sci.* **111,** 7980-7985 (2014).

**Acknowledgements**

We thank Paul Ginsparg, Jenny Hoffman, Steve Kivelson, Roger Melko, Andy Millis, Miles Stoudenmire, Kilian Weinberger and Jan Zaanen for helpful discussions and communications; A.M., and Y.Z. acknowledge support from DOE DE-SC0010313; Y.Z. acknowledge support from DOE DE-SC0018946; E.-A.K. and J.C.S.D. acknowledge support by the Cornell Center for Materials Research with funding from the NSF MRSEC program (DMR-1719875); E.K. and K.C. acknowledge support from the NSF Grant No. DMR-1609560. E.-A.K. and E.K. acknowledge support from the Kavli Institute for Theoretical Physics, supported in part by the NSF under Grant No. PHY-1748958, where initial discussions about the project took place. S.U. and H.E. acknowledge support from a Grant-in-aid for Scientific Research from the Ministry of Science and Education (Japan) and the Global Centers of Excellence Program for Japan Society for the Promotion of Science. K.F. and J.C.S.D. acknowledge support from the U.S. Department of Energy, Office of Basic Energy Sciences, under contract number DEAC02-98CH10886; S.D.E, M.H.H and J.C.S.D. acknowledge support from the Moore Foundation's EPiQS Initiative through Grant GBMF4544; J.C.S.D. acknowledges salary support from the European Research Council (ERC) under Award # DLV-788932.

**Author Contributions:** E.-A.K. and J.C.S.D. conceptualized the project. E.K., K.C., E.-A.K., and Y.Z. designed the machine learning strategy. Y.Z. implemented the ANN-based machine learning strategy. E.-A.K. and A.M. constructed the mathematical model for the training set and A.M. generated the training set. K.F. S.U. and H.E. synthesized and characterized the crystals studied. K.F, S.D.E and M.H.H carried out the experiments and image array data processing. E.-A.K. and J.C.S.D. supervised the investigation and wrote the paper with key contributions from K.F., Y. Z. and A.M. The manuscript reflects the contributions of all authors.

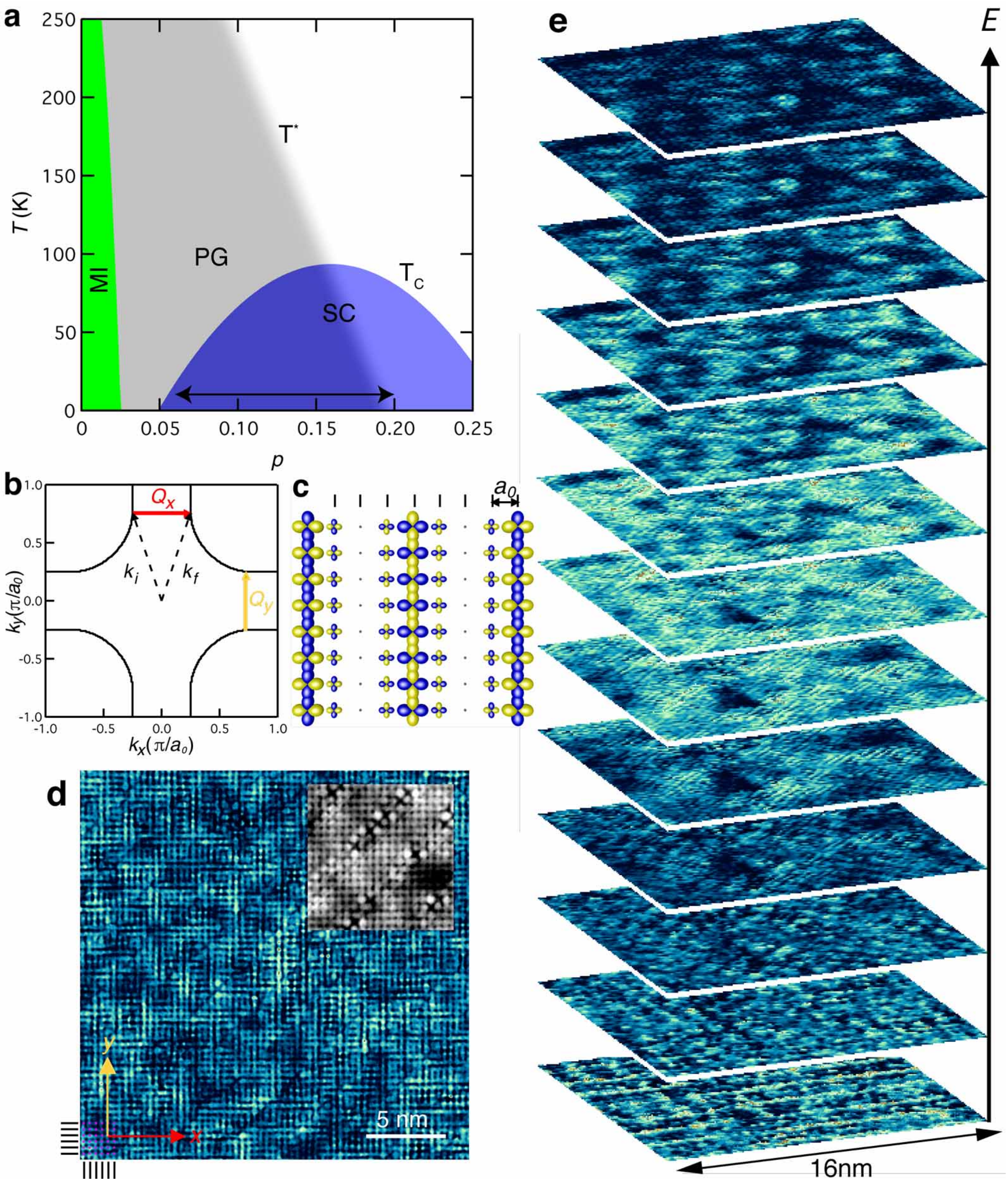

**Figure 1. Electronic quantum matter imaging in hole-doped $CuO_2$. a,** Schematic phase diagram of hole-doped $CuO_2$. At *p=0* a single electron is localized at each Cu site in a Mott insulator (MI) state. As holes are introduced (electrons removed) the MI disappears quickly. The high temperature superconductivity (SC) emerges at slightly higher p, reaching its maximum critical temperature $T_c$ near *p~0.16*. However, in the range *p<0.19* and up to temperatures $T^*$ an enigmatic phase of EQM, dubbed the pseudogap (PG) phase, is known to contain periodic charge density modulations of imprecise wavevector $\boldsymbol{Q}$. **b,** In the $CuO_2$ Brillouin zone, the Fermi surface is defined as the $\boldsymbol{k}$-space contour $\boldsymbol{k}(E=0)$ that separates the occupied from unoccupied electronic states, and its locus changes rapidly with changing carrier density $p$. Density wave (DW) states may then appear at a wavevector $\boldsymbol{Q}(\boldsymbol{k}_i(E=0)-\boldsymbol{k}_f(E=0))$ if the electron states $\boldsymbol{k}_i(E)$ and $\boldsymbol{k}_f(E)$ are "nested" (red and yellow arrows). **c,** Strongly correlated electrons may be fully localized in the Mott insulator phase, or self-organized into electronic liquid crystal states in $\boldsymbol{r}$-space. Schematically shown here is a simple example of a state with unidirectional charge density modulations in the $CuO_2$ plane**,** having wavelength $\lambda=4a_0$ or wavevector $\boldsymbol{Q}=\frac{2\pi}{a_0}(0.25,0)$ (Methods section 1). **d,** Typical 24.4nmX24.4nm SISTM image of electronic structure $R(\boldsymbol{r},E=150mV)$ from the $CuO_2$ plane of $Bi_2Sr_2CaCu_2O_8$ with *p=0.08* ($T_c$=45K). Complex spatial patterns, which to human visual perception look like highly disordered "tweed", dominate. The contrast with simple periodic arrangement of the simultaneously visualized atoms of the same crystal in the topograph (upper inset) is arresting. **e,** Typical image-array of simultaneously measured $Z(\boldsymbol{r},E)$ for *p=0.08*, each 16nmX16nm but at a different electron energy $E$, spanning the range $6meV<E<150meV$ in steps of 12 meV. Such arrays are the basic type of data-set for which efficient ML analysis and discovery techniques are required.

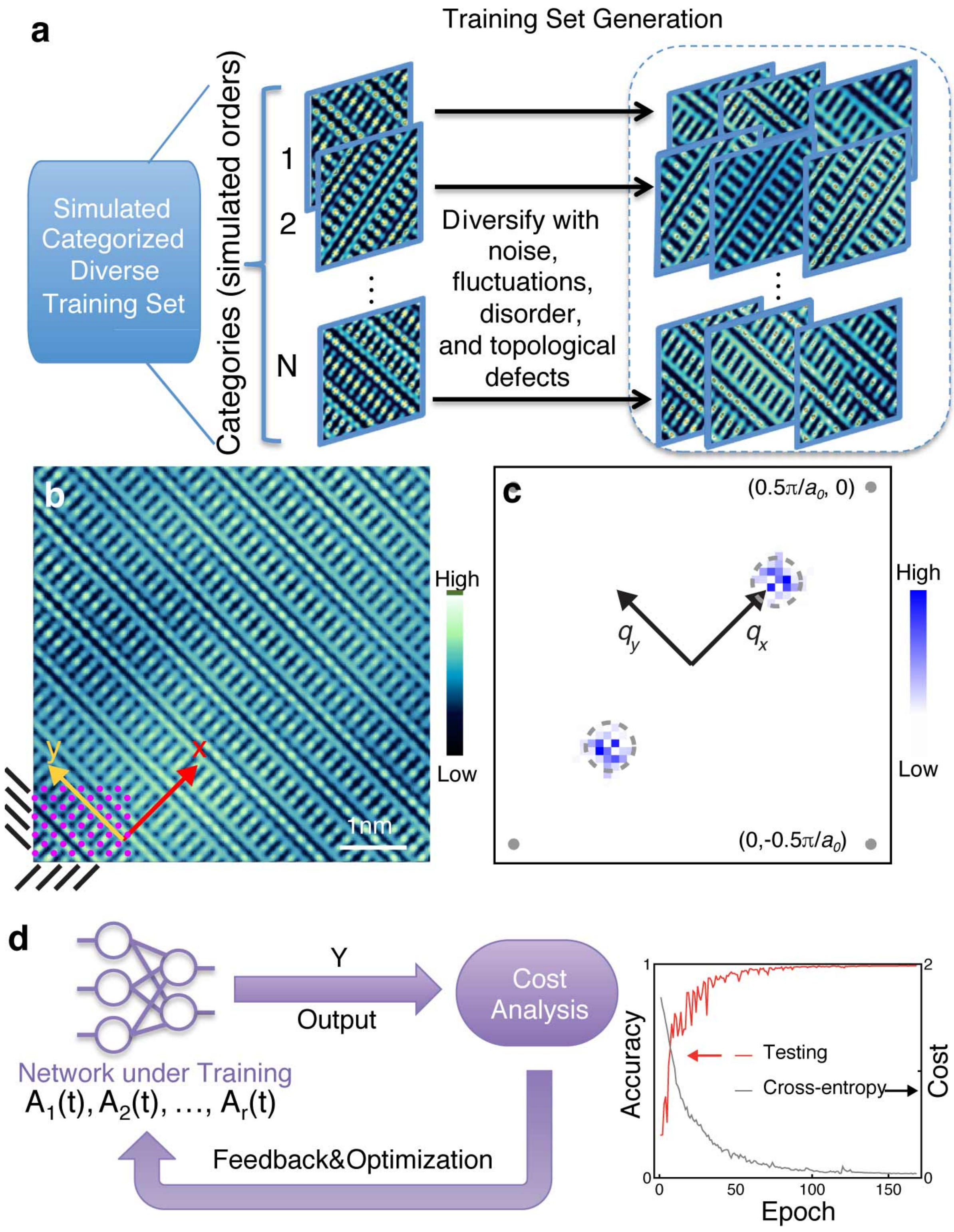
a
Training Set Generation
Simulated Categorized Diverse Training Set
Categories (simulated orders)
1
2
N
Diversify with noise, fluctuations, disorder, and topological defects
b
High
Low
y
x
1nm
c
(0.5π/a0, 0)
(0,-0.5π/a0)
qy
qx
High
Low
d
Network under Training
A1(t), A2(t), …, Ar(t)
Y
Output
Cost Analysis
Feedback&Optimization
Accuracy
Cost
Testing
Cross-entropy
Epoch

**Figure 2. Training ANN to identify broken-symmetry states in SISTM data. a,** The ANN array is trained to recognize a DW in electronic structure images (e.g. $Z(\boldsymbol{r}, E)$) representing different EQM states. A synthesized training-image set for the ANNs is obtained by appropriately diversifying pristine images of *4* distinct electronic ordered states. Each translational symmetry-breaking ordered state is labeled by a category $C = 1,2,3,4$ associated with its wavelength: $\lambda_C = 4.348a_0, 4a_0, 3.704a_0, 3.448a_0$ respectively. The training-images in each category are diversified by appropriate addition of noise, short correlation-length fluctuations in amplitude and phase, and topological defects. **b,** Example of a training-image in category $C$=2 which is a $d$-symmetry form factor (dFF) DW along x-axis with $\lambda = 4a_0$ within which smooth amplitude and phase fluctuations and randomized positions of topological defects (dislocations) have been added to simulate typical phenomena encountered in experimental EQM visualization (e.g. 1D). The full 516x516 pixel image contains 2x86x86 entire $CuO_2$ unit-cells with Cu-Cu distance of 6 pixels diagonally. **c,** The $d$-symmetry Fourier transform of **b.** Absence of a well-defined modulation wavevector $\boldsymbol{Q}$ within the modulations in **b** has been successfully simulated in the training-image as seen by the region of $\boldsymbol{q}$-space (grey dashed circle) within which strong variation in the amplitudes at different wavevectors occur. Grey dots are at $\boldsymbol{q} = \left[\frac{2\pi}{a_0}\right](\pm 0.5, 0); (0, \pm 0.5)$. **d,** Each ANN is trained by minimizing the cross-entropy cost function progressively through stochastic gradient descent and back propagation. The process of going through the entire set of shuffled training data, also known as an epoch, is repeated until the cross-entropy and accuracy saturate. The overall accuracy of the finalized ANNs on the synthesized data is generally over 99%.

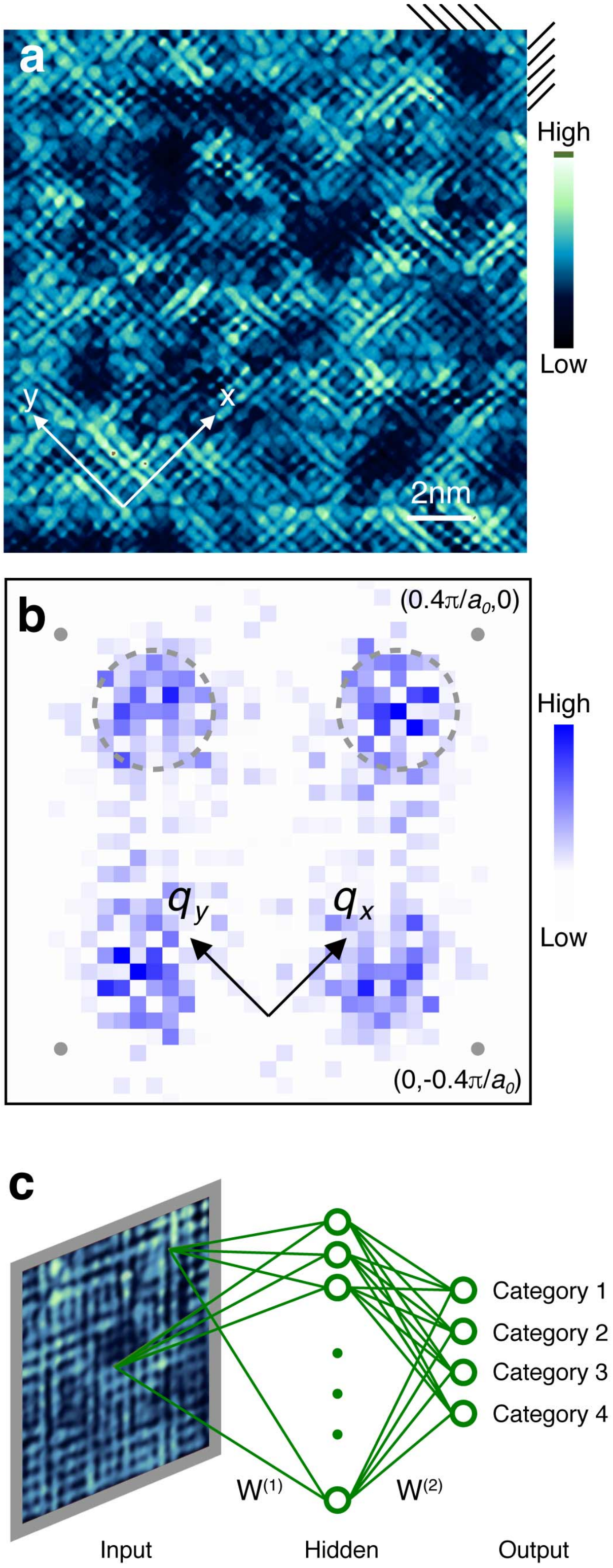

**Figure 3. ANN analysis of experimental EQM visualization data. a,** Typical measured 16nmX16nm $Z(\boldsymbol{r}, E = 84meV)$ image of $Bi_2Sr_2CaCu_2O_8$ with *p=0.08* ($T_c$=45K). The disorder and complexity of cuprate EQM are manifest. **b,** Typical measured $Z(\boldsymbol{q}, E = 84meV)$ image of $Bi_2Sr_2CaCu_2O_8$ with *p=0.08* ($T_c$=45K) being the d-symmetry Fourier transform of **a**. The disorder and complexity of EQM are equally vivid here in the broad and fluctuating peaks around $(Q_x \pm \delta Q_x, \delta Q_y)2\pi/a_0$ and $(\delta Q_x, Q_y \pm \delta Q_y)2\pi/a_0$ with $|\delta Q_x| = |\delta Q_y| \approx 0.2$. Grey dots are at the $(0.4,0); (0,0.4)2\pi/a_0$ points. **c,** Schematic of ANN analysis procedure for experimental $Z(\boldsymbol{r}, E)$ images: the successfully trained neural network with fixed parameters (weights $W^{(1)}$ and $W^{(2)}$ of the hidden layer and the output layer respectively and biases) is a classifier: It classifies each experimental image as belonging into one of the four categories. Neuron activation functions in our ANNs are taken to be the sigmoid function.

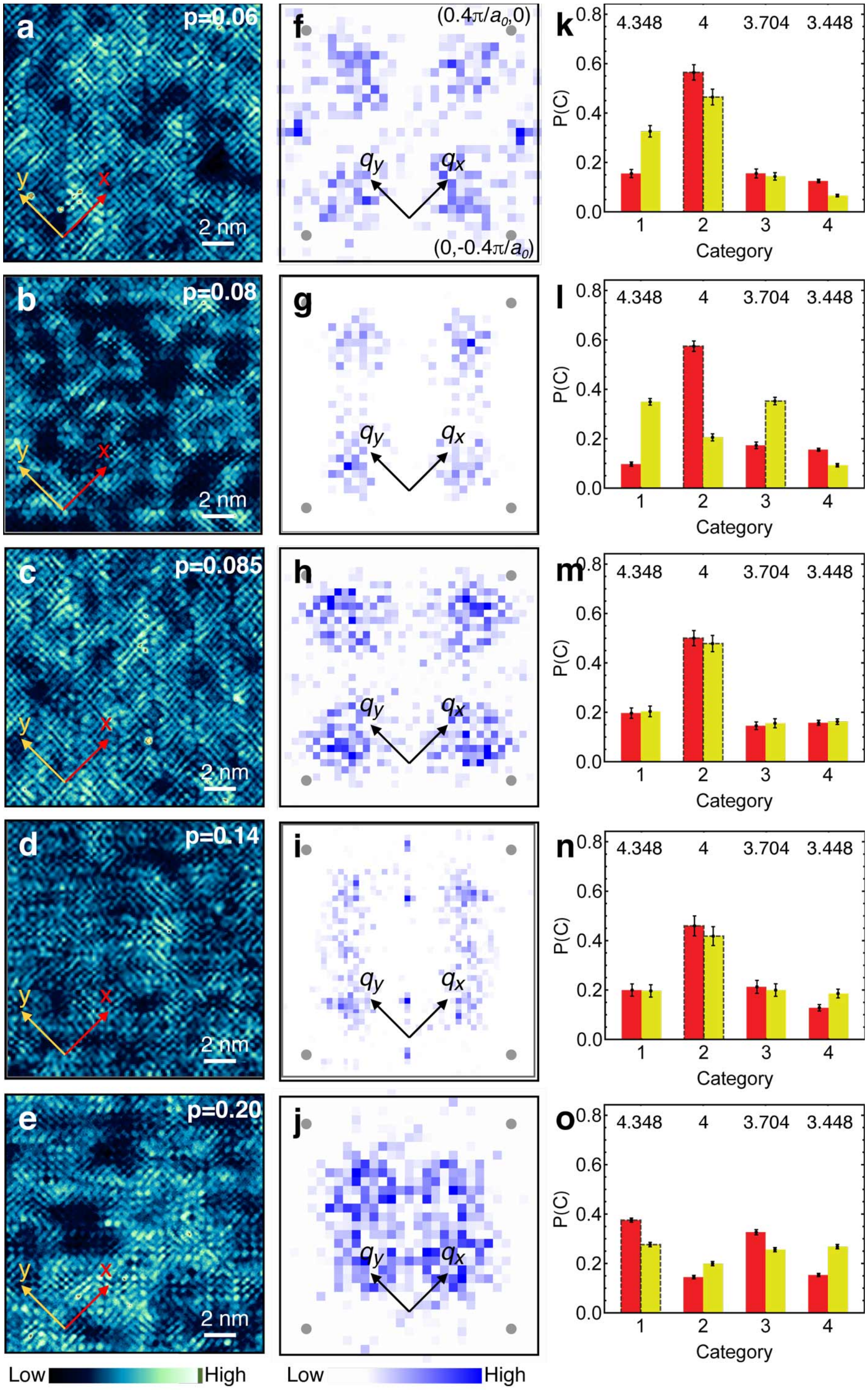

**Figure 4. ANN detection of broken-symmetry evolution with electron-density. a-e,** Measured 16nmX16nm $Z(\boldsymbol{r},E)$ images of $Bi_2Sr_2CaCu_2O_8$ in the range *p=0.06,0.08,0.085,0.14,0.20* ($T_c$(K)=20,45,50,74,82). Each image is measured at $E=\Delta_1(p)$ the pseudogap energy at that electron-density. Obviously disorder and complexity of cuprate EQM abound throughout this whole electron-density range (black double headed arrow in Fig. 1A). **f-j,** The $d$-symmetry Fourier transforms $Z(\boldsymbol{q},E)$ from **a-e**. The disorder and complexity of EQM are equally vivid as broad fluctuating peaks around $(Q_x \pm \delta Q_x, \delta Q_y)2\pi/a_0$ and $(\delta Q_x, Q_y \pm \delta Q_y)2\pi/a_0$. Grey dots are at the $(0.4,0);(0,0.4)2\pi/a_0$ points. **k-o,** Output categorization by 81 ANNs of the input data from **a-e.** Top row numbers: the category's fundamental wavelength, in units of $a_0$. We take statistics of independent assessment on the given experimental image by 81 ANN's that are independently trained to arrive at the probabilities *P(C)* of the image belonging to category *C*. The error bars mark the statistical spread (one standard deviation) of P(C) among 81 independently trained ANN's (see Methods). Since the training-images for ANNs are unidirectional, i.e., their pristine orders are along x-axis, categorization results for two modulation orientations X,Y (red and yellow bars) are obtained by inputting to ANNs the $Z(\boldsymbol{r},E)$ images and their 90-degree rotated versions, respectively.

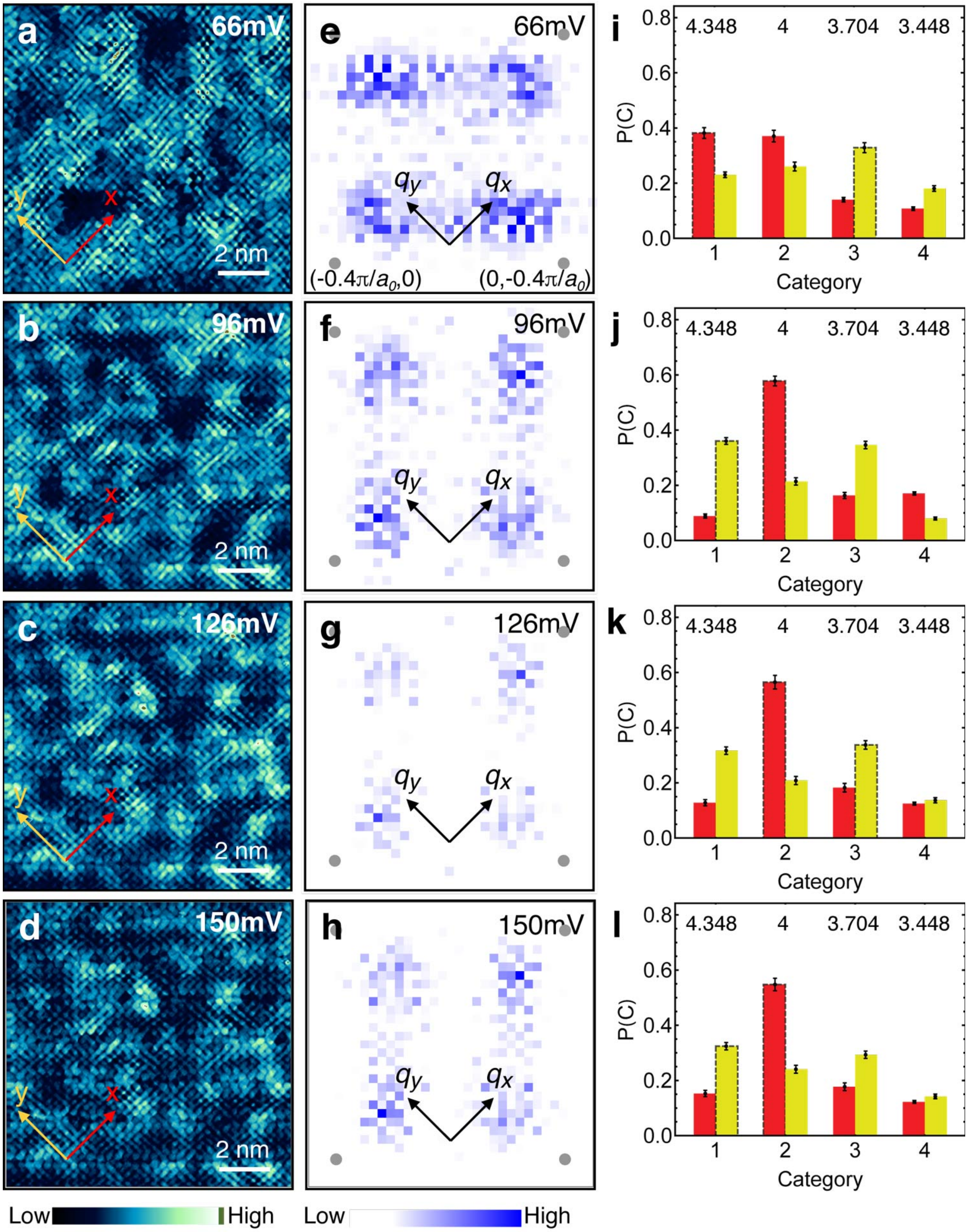

**Figure 5. ANN detection of broken-symmetry at different electron-energies. a-d,** Measured 16nmX16nm $Z(\boldsymbol{r},E)$ images of $Bi_2Sr_2CaCu_2O_8$ in a range of electron-energy *E=66,96,126,150 (meV) for p=0.08* ($T_c$(K)=45K). EQM complexity in the identical field of view, now evolves rapidly with electron-energy, a purely quantum mechanical effect. **e-h,** The *d*-symmetry Fourier transforms $Z(\boldsymbol{q},E)$ from **a-d**. The disorder and complexity of EQM are strong as seen in the broad fluctuating peaks around $(Q_x \pm \delta Q_x, \delta Q_y)2\pi/a_0$ and $(\delta Q_x, Q_y \pm \delta Q_y)2\pi/a_0$ but now $\delta Q_x, \delta Q_y$ evolve rapidly with electron-energy (another quantum mechanical effect). Grey dots occur at $(0.4,0); (0,0.4)2\pi/a_0$ points. **i-l,** Output categorization by 81 ANNs of the input data from **a-d**. Top row numbers: the category's fundamental wavelength, in units of $a_0$. We take statistics of independent assessment on the given experimental image by 81 ANN's that are independently trained to arrive at the probabilities *P(C)* of the image belonging to category *C*. The error bars mark the statistical spread (one standard deviation) of P(C) among 81 independently trained ANN's (see Methods). Categorization results for two modulation orientations X,Y (red and yellow bars) are obtained by inputting to ANNs the $Z(\boldsymbol{r},E)$ image-array and its 90-degree rotated version, respectively.

# METHODS

***1 Strong coupling density wave states.*** Real space (position based), strong coupling theories for carrier doped $CuO_2$ predict lattice commensurate, unidirectional, density waves in various electronic degrees of freedom. Among them are two candidate states that can both lead to $4a_0$-periodic modulations of the charge density and of the local density of states $N(\boldsymbol{r})$ with wavevector $\boldsymbol{Q} = (2\pi/4a_0, 0)$. First, a $4a_0$-periodic modulation in the charge density on the two oxygen sites $O_x$ and $O_y$ within each unit cell but with a relative phase $\pi$ between them. This is a d-symmetry form factor charge density wave existing as a fundamental ordered state. Second, an $8a_0$-periodic modulation of $d$-wave Cooper pair density can exist as a fundamental ordered state, and it induces a $4a_0$-periodic modulation in the charge density. These two distinct fundamental states are shown schematically in Extended Data Fig.2a,b respectively.

***2 Fourier Transform Analysis of EQM Images: Disorder and Information Loss.*** A Fourier transform of two-dimensional image data is a linear transformation of that data. All the information that was in the original image appears in the full, complex Fourier transform throughout reciprocal space. Importantly, when there are complicated local patterns or motifs of short-range order at atomic-scale in real space, that information gets spread over *all* of reciprocal space. This is because what is extremely local in real space becomes completely delocalized in reciprocal space. But, in the traditional mode of FT analysis, one typically picks a compact region in the reciprocal space to be of importance, because the intensity is peaked at that point. Crucially in this approach there is abundant information throughout reciprocal space away from the peak-intensity wavevector that has been discarded. For hole-doped $CuO_2$ the real-space electronic structure at atomic scale is uniquely complex (Fig. 1). For instance, one always finds that the STM image whose FT peak-intensity occurs away from Q=0.25 (see Extended Data Fig.3a), hosts vivid local motifs that are commensurate with the lattice (see Extended Data Fig.3b). Since any information local in position space gets spread over all reciprocal space, when one discards much of the data throughout reciprocal space crucial insights contained in atomic scale electronic-structure motifs, discommensurations and topological defects are all lost. On the other hand, because of the versatility of ANN to capture any function whatsoever,[33] the new ML approach allows one to impartially inspect the entirety of the data in each STM image with no loss of information. This is a key distinction between the traditional FT approach and the ML approach which impartially analyzes all the data throughout real space.

***3 Training image set generation.*** The diversification of synthetic images of a unidirectional DW to create a training image set (see Extended Data Fig.4) starts from d-wave and s-wave form factor (DFF and SFF) components, and includes (1) heterogeneity through independent amplitude and phase fluctuations and (2) topological defects or dislocations in DFF. For any of the C=1,2,3,4 categories with representative wavelength $\lambda_C$, the DFF ($I^{DFF}_{C,f,d}$) and SFF ($I^{SFF}_{f}$)form factor modulations with noise models were

$$I^{DFF}_{C,f,d}(x,y) = A_{DFF}\left[1+\varepsilon_A\, A_f(x,y)\right] A_d(x,y)\, Cos\left(2\pi x/\lambda_C + \varepsilon_\varphi \varphi_f(x,y) + \varphi_d(x,y) + \varphi_{DFF}\right),$$

$$I^{SFF}_{f}(x,y) = A_{SFF}\left[1+\varepsilon_A\, A_f(x,y)\right] Cos\left(\varepsilon_\varphi \varphi_f(x,y) + \varphi_{SFF}\right), \quad \text{(S1)}$$

with overall constants $A_{DFF}=1$, $A_{SFF}=0.5$ and phase offsets $\varphi_{DFF} = \pi/4, \varphi_{SFF} = 0$. Here the amplitude field $A_f(x,y)$ and the phase field $\varphi_f(x,y)$ capture smooth fluctuations (different random realizations in $I^{DFF}_{C,f,d}(x,y)$ and $I^{SFF}_{f}(x,y)$), and $A_d(x,y)$, $\varphi_d(x,y)$ capture dislocation defects. For each category, we generate different realizations labeled by *f* and *d.* For each realization $f$ the $A_f(x,y)$ field is two-dimensional Gaussian fluctuation field with spatial length scale $\xi_A$=8$a$, normalized between (-1) and 1, while $\varphi_f(x,y)$ is two-dimensional Gaussian fluctuation field with the same spatial lengthscale $\xi_\varphi$=8$a$, normalized between -π and π. The values of correlation lengthscales $\xi_A$, $\xi_\varphi$ are motivated by a simple analysis of an SI-STM $Z(\boldsymbol{q},E)$ Fourier transform (Fig. 3). The strengths of amplitude and phase fluctuations $\varepsilon_A$=0.8, $\varepsilon_\varphi$=0.5 are also chosen to produce images in rough consistency with a typical $Z(\boldsymbol{r}, E)$. In each image, there are $n_d$=2 dislocations at random positions $\boldsymbol{x}_i=(x_i,y_i)$, $i$=1…$n_d$, with windings $w_i$=±2π and total winding 0. The total dislocation-contributed fields are:

$$A_d(\boldsymbol{x}) = \prod_{i=1}^{n_d} (1 - exp(-|\boldsymbol{x} - \boldsymbol{x}_i|/\xi_d))$$

$$\varphi_d(x,y) = \sum_{j=1}^{n_d} Arg\left[sgn(w_j)\left(x - x_j\right) + i\left(y - y_j\right)\right],$$

where the amplitude recovery length is $\xi_d=a$, motivated by $Z(\boldsymbol{r}, E)$.

Then the training set for each category C combines the different form factor components into image intensity at pixel position *(x,y)* in units of $a$ through

$$I_C(x,y) = I_{C,DFF}(x,y) * D(x,y) + I_{SFF}(x,y) * S(x,y),$$

using atomic masks: the SFF mask $S(x,y)$ is a sum of two-dimensional Gaussians with maxima equal to one and spatial widths equal to *0.35a*, each located at a *Cu* atom position (*x,y* integer), while the DFF mask $D(x,y)$ is a sum of positive Gaussians at locations of $O_x$ and negative ones at $O_y$'s. The total intensity $I_C(x,y)$ of all simulated images is normalized to take values between 0 and 1. All simulated images have 6 pixels per nearest *Cu-Cu* distance $a$, and contain 2x86x86 unit-cells, for the total size of 516x516pixels.

***4 Configuration of Artificial Neural Network (ANN).*** In a feed-forward fully-connected artificial neural network, the neurons form a layered structure and the output of each neuron is sent to all the neurons in the subsequent layer. Each neuron assesses all the inputs with a series of weights **w**, and an additive constant $b$ known as the bias, and determines the output through a non-linear transformation $f(\boldsymbol{w} \cdot \boldsymbol{x} + b)$, called the activation function. The bias $b$ and the weights $\boldsymbol{w}$, are the parameters of the ANN and adjusted during the training. The activation function usually takes the form of the sigmoid function or the rectified linear unit, see the inset of Extended Data Fig.5a. We also use a softmax function $\sigma(\boldsymbol{x})_j = e^{x_j} / \sum_j e^{x_j}$ for the output layer to normalize the output and allow a probabilistic interpretation for the different categories.

For supervised machine learning, we divide the data set into a training set containing 90% of the images and the rest 10% for unbiased validation, speed control, and overfitting detection during the training. The weights and biases of the ANN are optimized using stochastic gradient descent to minimize the cross-entropy cost function:

$$C = \frac{1}{N}\sum_{x}\sum_{i=1}^{4}[y_i ln(\sigma_i) + (1 - y_i) ln(1 - \sigma_i)],$$

where $y_i$ and $\sigma_i$ are, respectively, the desired output consistent with the label and the actual ANN output for each of the input image data $\boldsymbol{x}$. We use a batch size of 50, and L2 regularization to avoid overfitting. We include 50 neurons in the hidden layer and choose the sigmoid function as the neuron activation function unless stated otherwise. In Extended Data Fig.5a we show examples of the cost function as well as the accuracy on the validation data set for both choices of the sigmoid and the ReLU activation functions during the training. Extended Data Fig.5b shows the achieved accuracy and cross-entropy cost after 25 epochs as a function of the number of neurons in the single hidden layer. We have trained 81 ANNs with random initial conditions by using a stochastic training process. The outputs of the finalized ANNs are robust and quantitatively consistent with each other. Our results in the main text show the average and standard deviations from all 81 ANNs.

To verify that our results are robust against changes to the architecture of the ANN, we have trained 6 ANNs with 100 neurons in the single hidden layer, and 6 ANNs with two hidden layers, and we found that the results agree with each other within error bars.

Because they are drawn from a historic image-array archive not designed for ML based studies, the SI-STM image-arrays $Z(\boldsymbol{r}, E)$ vary in spatial resolution from sample to sample from 1.7 to 11.5 pixels per *a,* the average *Cu-Cu* distance. The number of $CuO_2$ unit-cells in experimental images also varies from 2x55x55 to 2x175x175. The *Cu* and $O_{x,y}$ atom positions, registered from the topograph, show random distortions of the lattice due to the STM tip drift effect (Extended Data Fig.6a).

To correct for the drift and standardize all the $Z(\boldsymbol{r}, E)$, we prepare each $Z(\boldsymbol{r}, E)$: (1) using interpolation we map the $Z(\boldsymbol{r}, E)$ to the resulting input image, in a way that each topographic atom position maps onto a position in a perfect atomic lattice with *Cu*-*Cu* distance of $a$=6pixels (see Extended data Fig.6b,c), which corrects both the drift effect and standardizes the spatial resolution; (2) we crop or tile the image to size 516x516pixels; (3) to study the degree of unidirectionality, for each input image we create a copy rotated by $90^{o}$, since the training images have modulations only along X direction for simplicity and clarity. An example *Mathematica* notebook file for data preparation is available. Extended Data Fig.7 shows the $Z(\boldsymbol{r}, E)$ and prepared input data at different dopings of $Bi_2Sr_2CaCu_2O_8$. It should be noted that the results are reliable only if the test data lie reasonably consistently within the input space given by the synthetic training sets.

***5 Validation and Benchmarking.*** To assess the discriminatory power of ANNs' categorization, we study obvious modulations in two experimental images (Extended Data Fig.8): (1) Topograph of $Bi_2Sr_2CaCu_2O_8$, which has no human-discernible modulation except for the *Cu* atomic lattice (an SFF at *Q=0*); (2) $Z(\boldsymbol{r}, E)$ of NCCOC, with obvious commensurate period $4a_0$ modulations, apparent in a DFF Fourier transform. The ANNs' categorization is in full accord.

We also checked the robustness of our approach against existence of $Bi_2Sr_2CaCu_2O_8$ superlattice modulations. The assessment of the ANN's were independent of existence or absence (data with superlattice modulation removed from the FT) of the superlattice modulations.

We further tested the robustness of the ANN decisions against change in the disorder model. For this we trained a new ANN with the training set generated with different disorder parameters. Specifically, we decreased the amplitude fluctuation intensity $\varepsilon_A$ by 13%, and phase fluctuation intensity $\varepsilon_\varphi$ by 20%, while making the disorder profiles vary more rapidly in space by decreasing the correlation lengths $\xi_A$, $\xi_\varphi$ by 6%.

Repeating the assessment of experimental data shown in Fig.4k,l,m,o and Extended Data Fig.1a with the new ANN, we find the results remain unchanged. This is shown through the comparison between the reprint of Fig.4k,l,m,o and Extended Data Fig.1a here as Extended Data Fig.9a-e respectively and the output from the ANN trained with the new disorder model as Extended Data Fig.9f-j. Robust observations are 1) preference for the commensurate period $4a_0$ for systems with $0.06<p<0.14$ (Extended Data Fig.9a-d, and f-i) and complete confusion over different candidate categories for $p$=0.2 (Extended Data Fig.9e and j). The energy dependence comparison between the ANN's assessments in the main text (Extended Data Fig.1a or Extended Data Fig.9e) and the assessments of the ANN trained with the altered disorder model (Extended Data Fig.9j) shows that the tie between the onset of preference for the commensurate period $4a_0$ and the nematicity at the pseudogap energy scale is equally robust against variations in the disorder model used to train ANN's.

***6 Discommensurations and Maximum Intensity Wavevector.*** The Fourier transform(FT) based linear analysis of equivalent data in Ref.28 was carried out using the fact that the power spectral density is not smoothly distributed (Extended Data Fig.10a,b, reproduced from the SI of Ref.28.). We had introduced the concept of demodulation residue (DR), using

$$\boldsymbol{R}_{\boldsymbol{q}}^{\alpha}[\psi] \equiv \int \frac{d^2\boldsymbol{x}}{L^2} Re\left[\Psi_{\boldsymbol{q}}{}^*(\boldsymbol{x})(-i\partial_\alpha)\Psi_{\boldsymbol{q}}(\boldsymbol{x})\right],$$

$\alpha = x, y$, which measures the phase fitness of the $\boldsymbol{q}$-modulation in spatial pattern $\psi(\mathbf{r})$ through filtered FT:

$$\Psi_{\boldsymbol{q}}(\boldsymbol{k}) = exp\left(-\frac{(\boldsymbol{k}-\boldsymbol{q})^2}{2\Lambda^2}\right) exp(-i\,\boldsymbol{q}\cdot\boldsymbol{x})\,\tilde{\psi}(\boldsymbol{q}+\boldsymbol{k}), \quad \text{(S2)}$$

where $\tilde{\psi}(\boldsymbol{k})$ is the FT of the data. By minimizing the DR, $R_{\boldsymbol{q}}[\psi] \equiv \sqrt{\boldsymbol{R}_{\boldsymbol{q}}^{x}[\psi]^2 + \boldsymbol{R}_{\boldsymbol{q}}^{y}[\psi]^2}$, for a given modulation while considering different $\boldsymbol{q}$-modulations, we showed that one can obtain the phase averaged wave vector $\bar{\boldsymbol{Q}}$ of DW modulations. Within the limits of Fourier transform, which is a linear basis transform, this approach was an advancement in dealing with situations when the amplitude does not show well defined peaks, due to severe disorder.

However, there are limitations in this approach because FT is a linear transformation of basis and is useful when the desired phenomenon has sharp features in the new basis: the wavevector basis. However, when there are randomly placed, highly disordered, patches of a real-space DW pattern with sprinkles of topological defects, Fourier transform based methods perform very poorly. Obviously, one would not attempt a Fourier transform in trying to recognize human faces in an image for precisely this reason. The limitation of the FT-based methods is evident in that, even when a modulation pattern consists of commensurate period $4a_0$ modulation ($Q_0=2\pi/4a_0$) everywhere except for a sequence of discommensurations (phase slips in commensurate modulation pattern), the $R_{\boldsymbol{q}}[\psi]$ minimization (as well as the FT amplitude maximization) incorrectly identifies an apparent period of $\bar{Q}=0.3*2\pi/a_0$ (Extended Data Fig.10e). Although in Ref.28 the DR minimization yielded $\bar{Q} = 2\pi/4a_0$ for pseudogap energy data (single data set for each doping) for various dopings, this depended critically on human visual inspection to identify commensurate patches in Fig. S6B of Ref.28 (see also Extended Data Fig.3 here). Furthermore, the DR based approach therein averaged over topological defects (dislocations) ignoring their role. Finally, the DR based approach required manual choice of Fourier cutoff ($\Lambda$ in Eq.S2) again based on human visual inspection of the data. Hence the entire process is time consuming and high-level human labor intensive and fraught with human perceptual bias. It is therefore not possible to study the largest SISTM image-arrays with this FT approach in any consistent way, rendering it impossible to inspect the complete electron-density and electron-energy dependence of the largest EQM image-array archives.

The ANN-based approach we introduce in the main text is far more powerful,

efficient and general. It does not rely on arbitrary choices such as cut-off $\Lambda$, or on visual selection of Fourier regions of interest, and is not tied to any basis. The ANN is inherently non-linear and an ANN with sufficient number of neurons can express/detect any function.[33] Due to the versatility of ANN's, our ANN-based approach allows us to rapidly analyze a complete image-array data set in its entirety, without any ad-hoc Fourier filtering or selection. Hence the ANN approach is quite unbiased. Moreover, once the ANN's are trained, the automatic assessment of new data set takes minutes, allowing for a high-throughput analysis. It is this efficiency that allowed discovery of the connection between nematic state and commensurate density wave state, both setting in at the pseudogap energy scale (Extended Data Fig.1).

## Methods References

33 Cybenko, G. Approximation by superposition of a sigmoidal function. *Math. Control Signals Systems* **2**, 303—314 (1989).

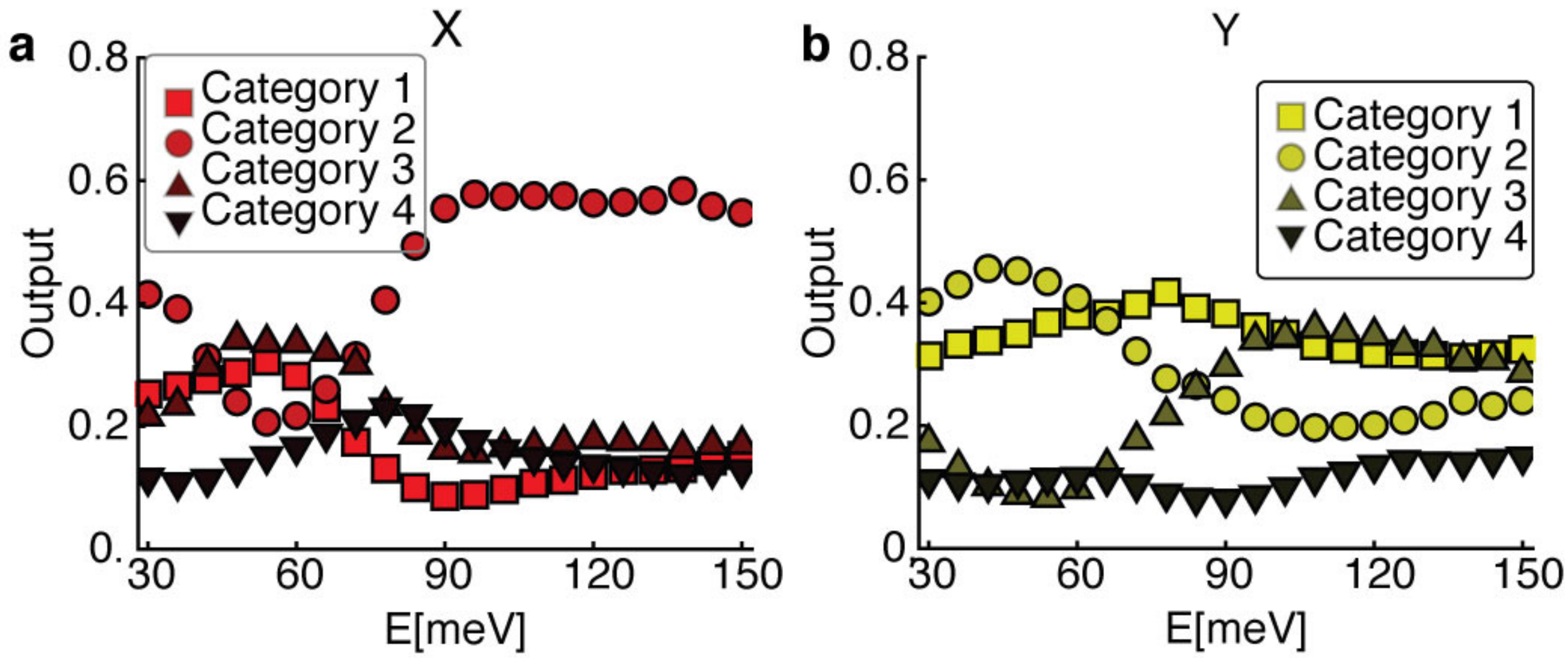

**Extended Data Fig. 1 ANN detection of unidirectionality at different electron-energies. a, b,** Output categorization by 81 ANNs of the 16nmX16nm $Z(\boldsymbol{r}, E)$ images of $Bi_2Sr_2CaCu_2O_8$ in a range of electron-energy *E=30...150 (meV)* in steps of 6 meV for *p=0.08* ($T_c$(K)=45K). Plot symbols are larger than the statistical spread (one standard deviation) of the ANN outputs, as estimated from our ensemble of 81 ANN realizations (see Methods). **a,** Output for modulation orientation X is obtained by inputting to ANNs the $Z(\boldsymbol{r}, E)$ image-array. **b,** Output for modulation orientation Y is obtained by inputting to ANNs the 90-degree rotated versions of the $Z(\boldsymbol{r}, E)$ used for **a**.

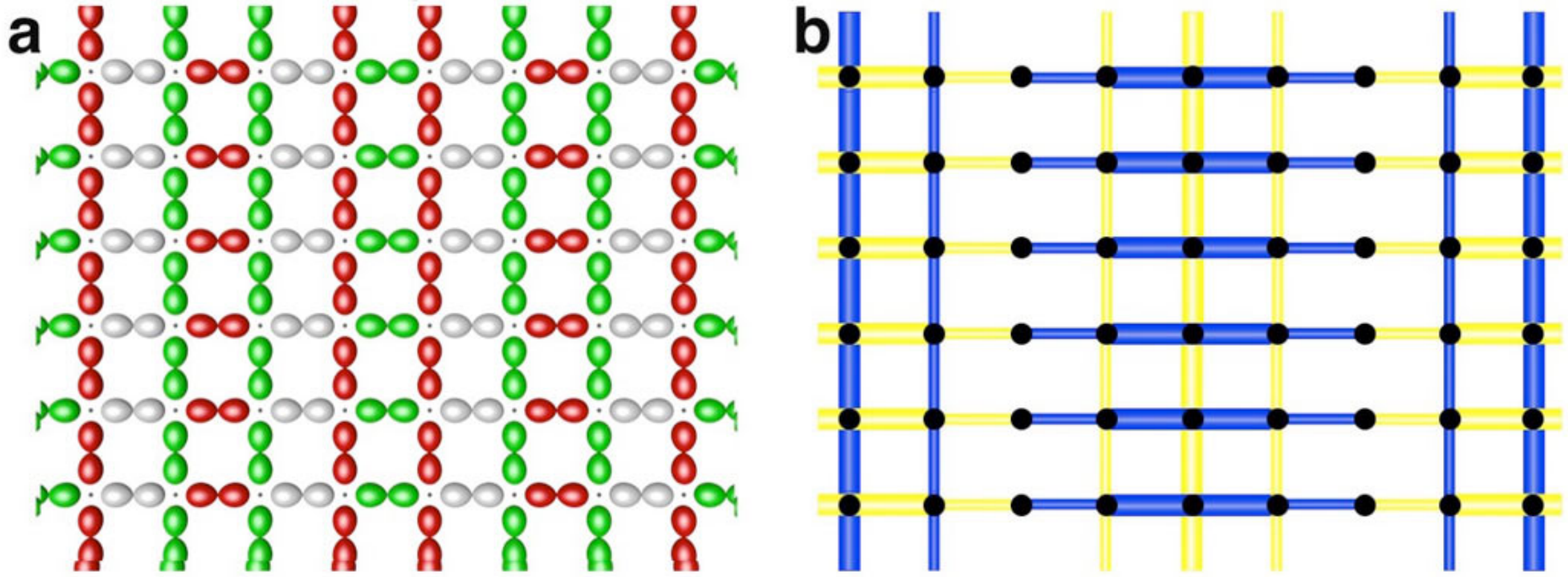

**Extended Data Fig. 2 Schematic image of density waves arising from strong coupling position-based theories in *$CuO_2$* plane. a,** The d-symmetry $4a_0$ charge density wave. The charge density at $O_x$ site is modulating with four-unit-cell periodicity along horizontal direction, and similarly for that at $O_y$ but out of phase by π (*d*-symmetry). *Cu* locations are marked by small dots. **b,** The $8a_0$ pair density wave state. The *d*-wave Cooper pair density is modulated with eight-unit-cell periodicity along horizontal direction. Such modulation in Cooper pair density can cause $4a_0$-period modulation in the local density of states *N(r)*.

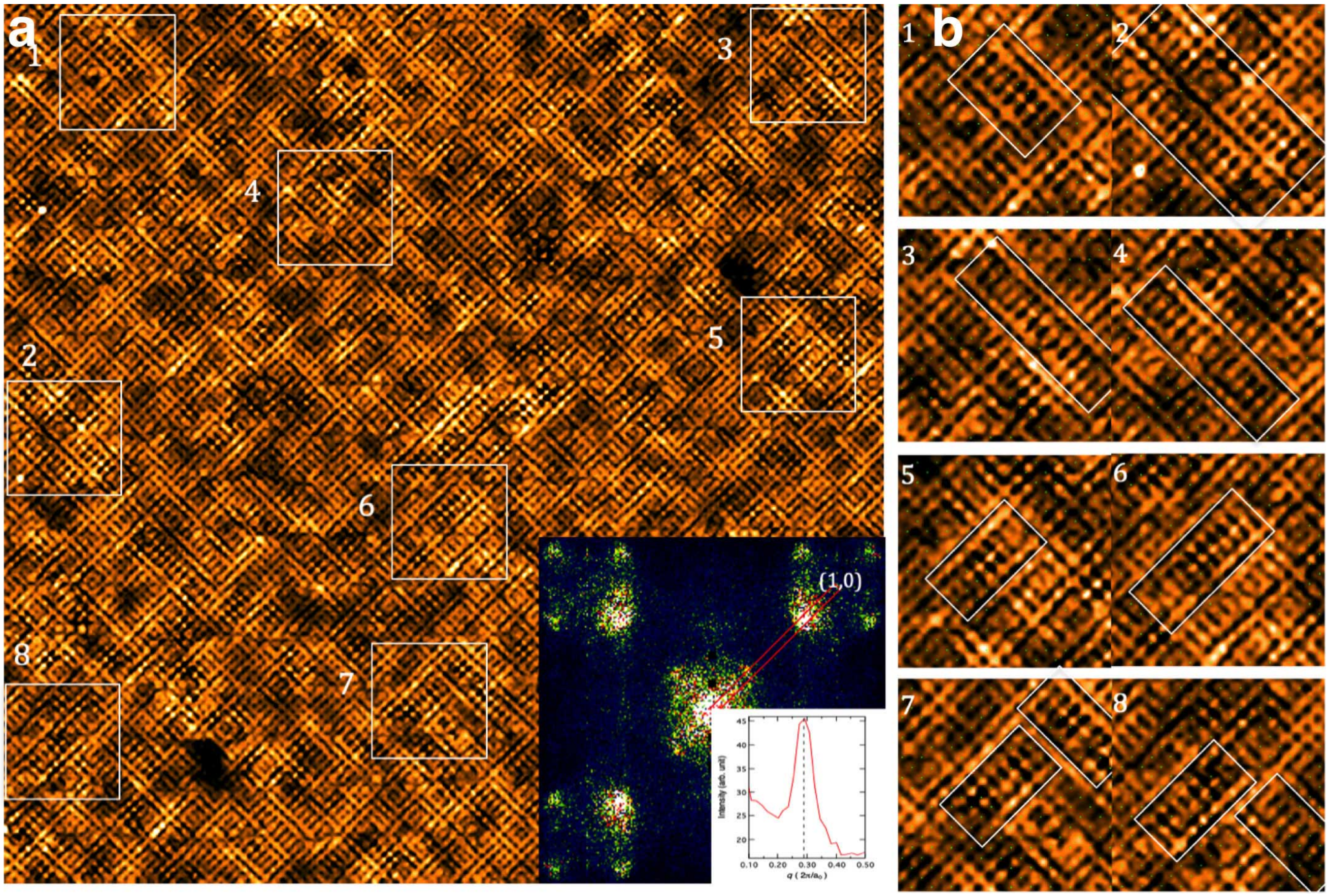

**Extended Data Fig. 3 Local commensurate motifs in STM images. a,** Large field of view, high precision, STM image of electronic structure integrated to E=100meV in $Bi_2Sr_2CaCu_2O_8$ with p~0.08. Inset shows the power spectral density FT while lower-right inset shows that data plotted along a line from 0.1 to 0.5 in units of $2\pi/a_0$. Clearly, the maximum intensity peak occurs at <Q>=0.28. **b,** Within each of the eight 6.5nm-square regions taken from **a** there are many commensurate, unidirectional $4a_0$ electronic-structure motifs (inside white rectangles). The Cu sites, independently determined from topographic imaging, are shown as fine dots.

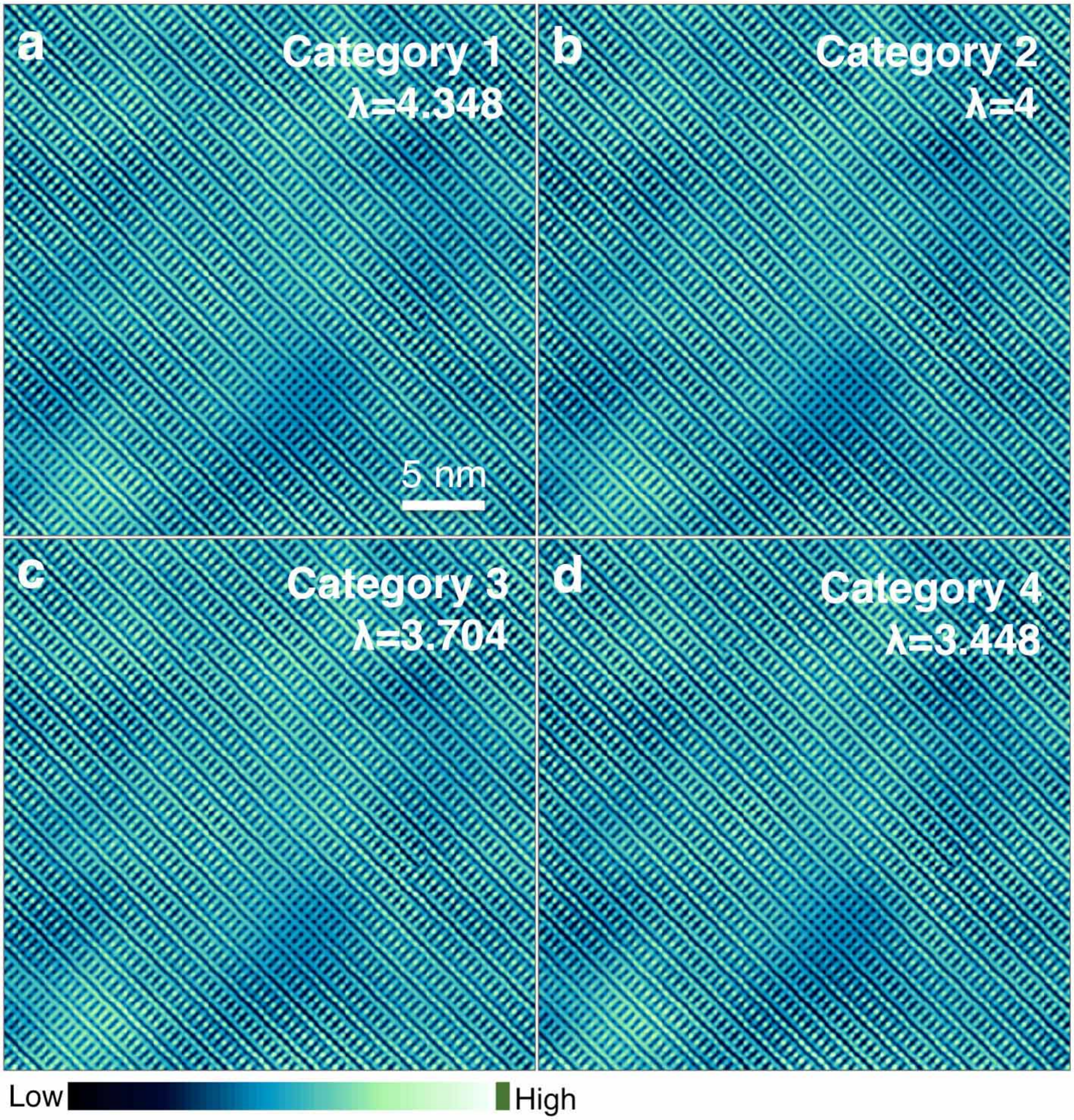

**Extended Data Fig. 4 Categories defined by electronic orders.** Example images from the simulated training set, from category *C=1* in **a**, *C=2* in **b**, *C=3* in **c**, *C=4* in **d**, defined by d-wave form factor unidirectional modulation with wavelengths $\lambda_C = 4.348a_0, 4a_0, 3.704a_0, 3.448a_0$, respectively. The $CuO_2$ unit-cell size is $a_0$=6pixels, diagonally.

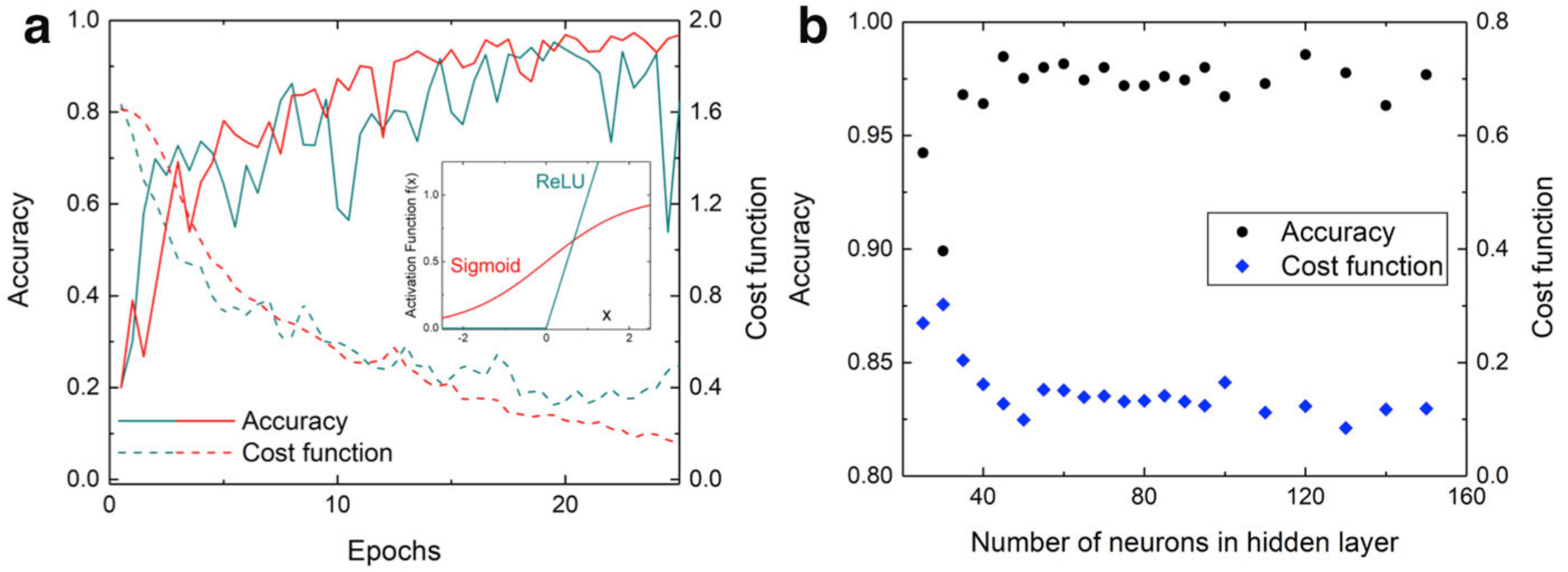

**Extended Data Fig. 5 Artificial Neural Network(ANN) training and testing. a,** Examples of the accuracy of the ANN outputs for the independent validation data set and the cross-entropy cost function is compared over different neuron activation functions during the initial training processes. The inset illustrates the non-linear activation functions - the sigmoid function and the rectified linear unit. **b,** Examples of the accuracy and the cross-entropy cost versus the number of neurons in the single hidden layer after 25 epochs of training.

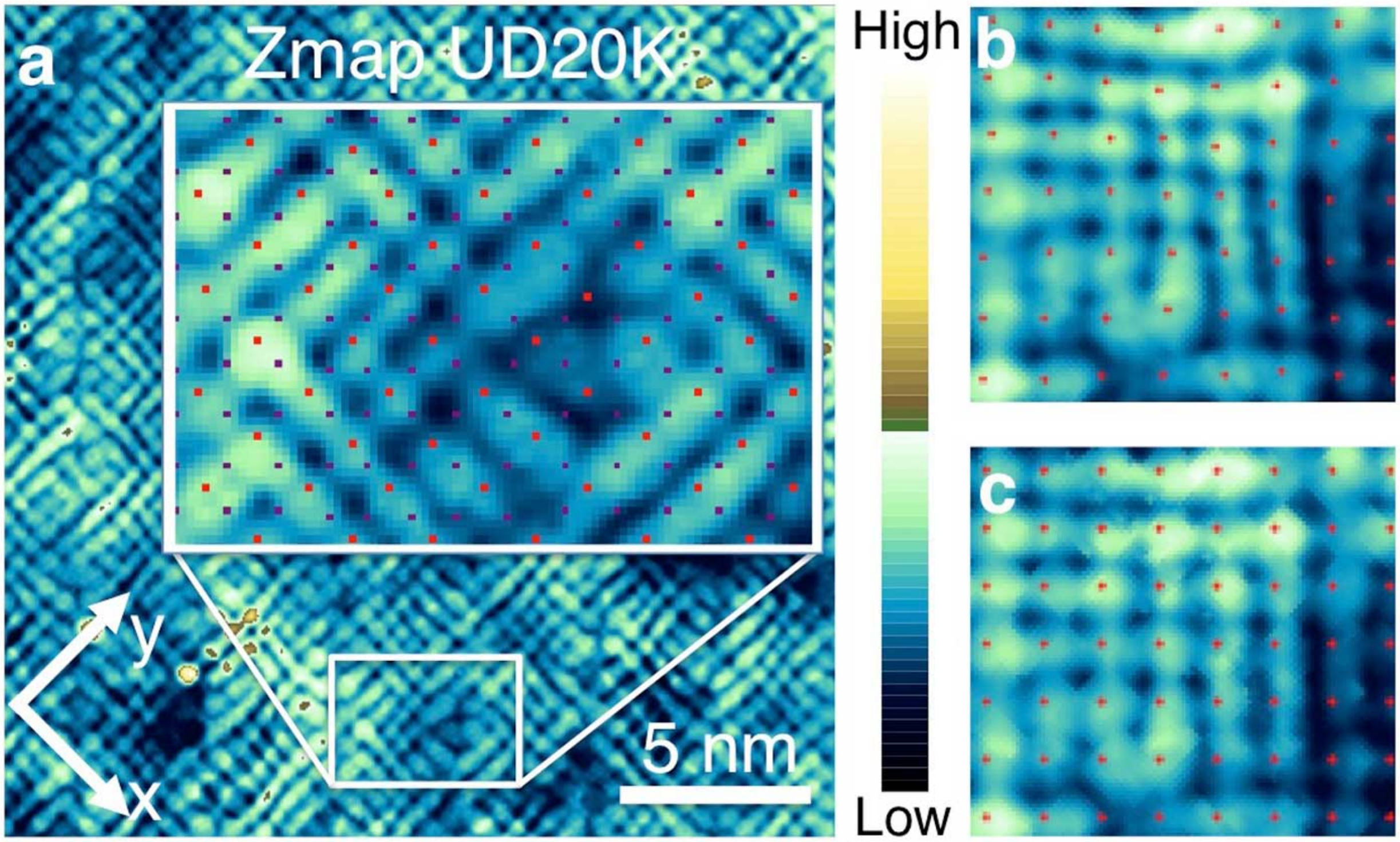

**Extended Data Fig. 6 Experimental SI-STM images. a,** Example $Z(\boldsymbol{r}, E)$ of underdoped $Bi_2Sr_2CaCu_2O_8$ with hole density $p=0.06$ ($T_c$(K)=20). The inset is a zoom-in with marked atom positions determined from topograph (Cu: red/light, O: purple/dark). **b.** A small region of **a**. **c,** The standardized version of **b** (see Methods).

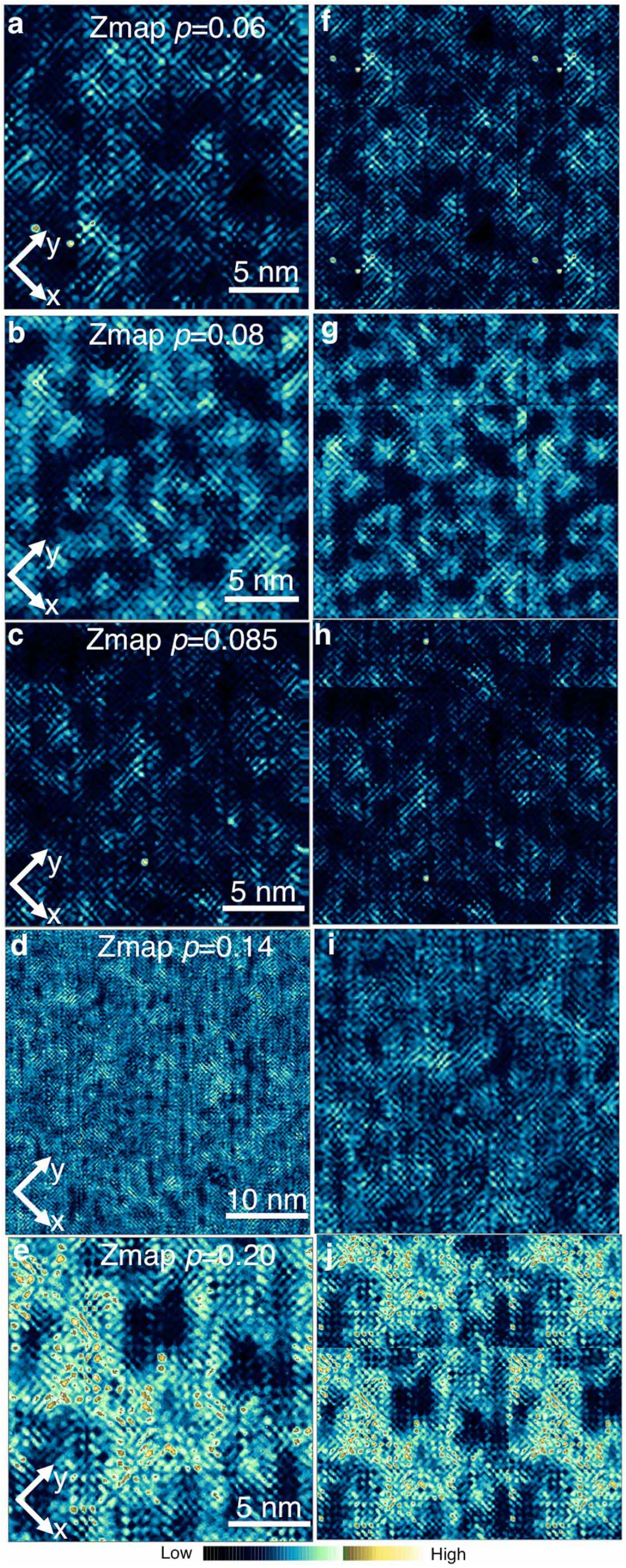

**Extended Data Fig. 7 Experimental SI-STM images as input for categorization. a,** The $Z(\boldsymbol{r}, E)$ of underdoped $Bi_2Sr_2CaCu_2O_8$ with *p=0.06* ($T_c$(K)=20) at energy $E$=$\Delta_1$ (see main text). **f,** The 516x516 pixel (2x86x86 $CuO_2$ unit-cells) input data from **a** (see Methods). **(b,g), (c,h), (d,i) and (e,j)** The same as the pair **(a,f)** but for, respectively, underdoped $Bi_2Sr_2CaCu_2O_8$ with *p=0.08* ($T_c$(K)=45), underdoped $Bi_2Sr_2CaCu_2O_8$ with *p=0.085* ($T_c$(K)=50), underdoped $Bi_2Sr_2CaCu_2O_8$ with *p=0.14* ($T_c$(K)=74), and overdoped $Bi_2Sr_2CaCu_2O_8$ with *p=0.20* ($T_c$(K)=82). Too small images are tiled, with unit-cells intact at the tiling boundary, while too large images are cropped.

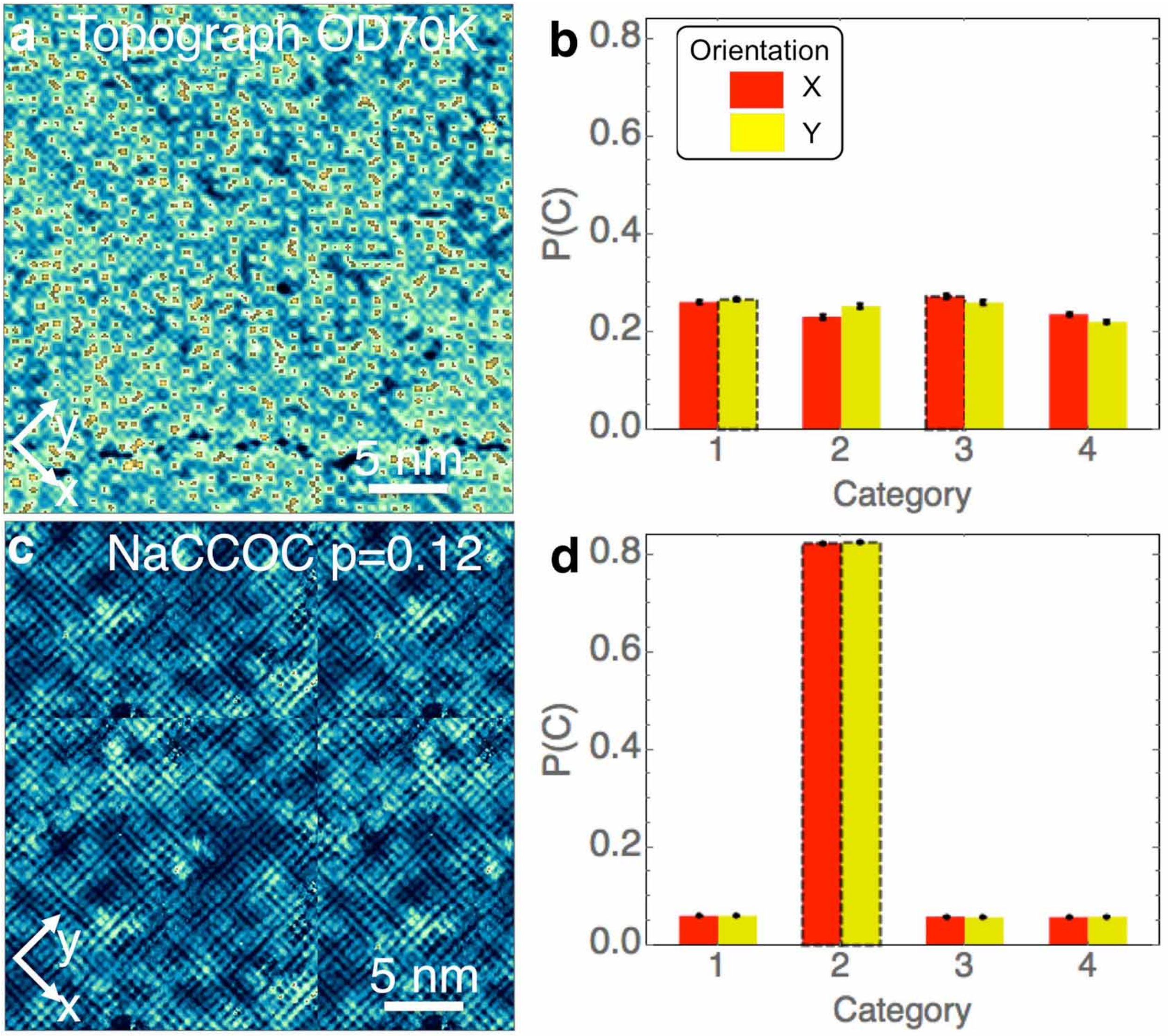

**Extended Data Fig. 8 Benchmarking categorization using experimental images. a,** Input data for the topograph of overdoped $Bi_2Sr_2CaCu_2O_8$ with *p=0.22* ($T_c$(K)=70). **b,** Output categorization by 81 ANNs of **a**, showing absence of translation-breaking signal. Results for two modulation orientations X,Y are obtained by inputting to ANNs the image in **a** and its 90-degree rotated version, respectively (see Methods). **c,** The input data for $Z(\boldsymbol{r}, E)$ of NCCOC at doping *p=0.12* at *E*=150meV. **d,** Output categorization by 81 ANNs of **c**, showing commensurate modulations (category 2).

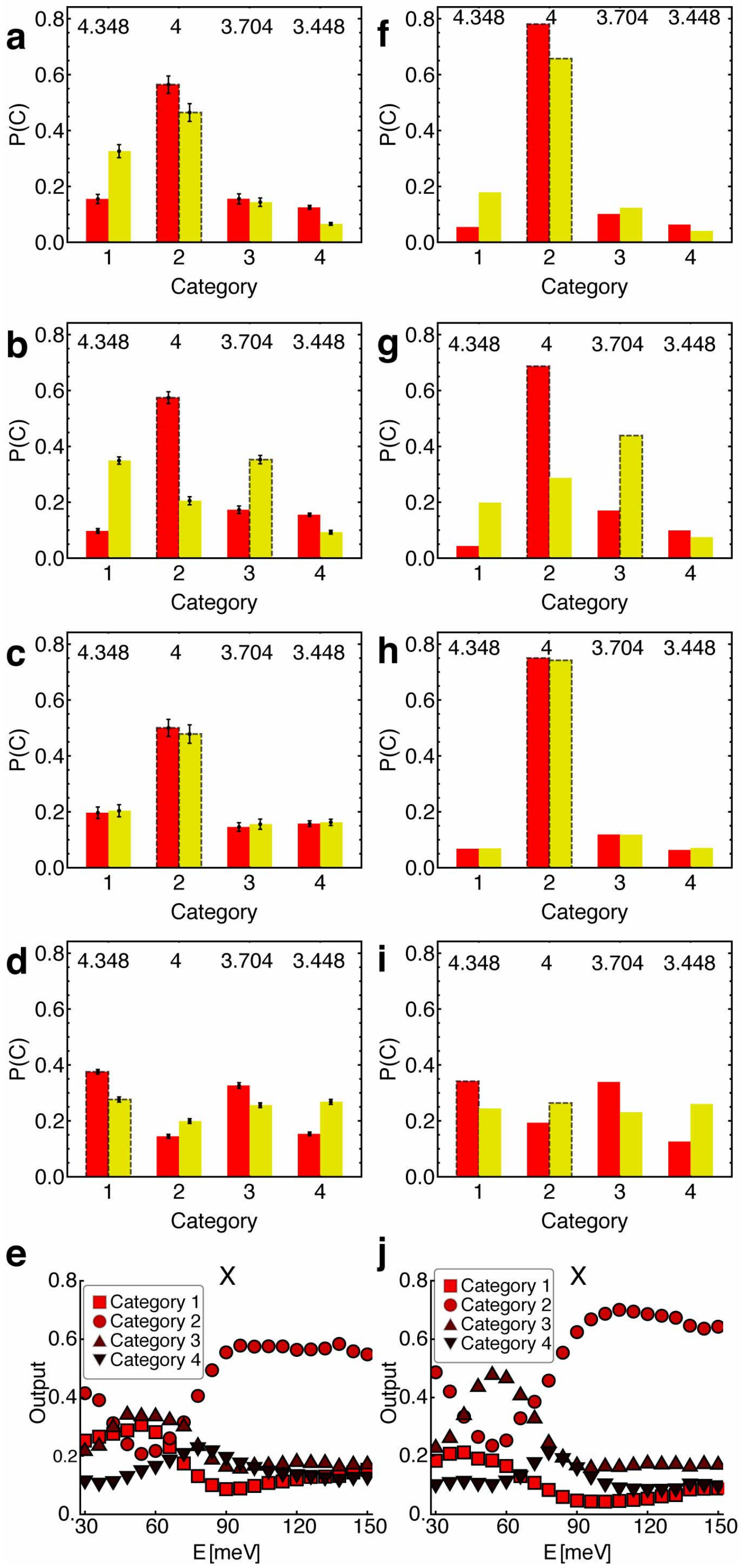

**Extended Data Fig. 9 Categorization is robust to changes in training set parameters.** **a-d,** Output categorizations of main Figure 4k,l,m,o showing evolution with hole doping. **e,** Output categorizations of Extended Data Figure 1a showing evolution with electron energy. **f-j,** Categorizations of the same inputs as for **a-e,** respectively, but obtained from output of a single ANN trained using a different training set (see Methods).

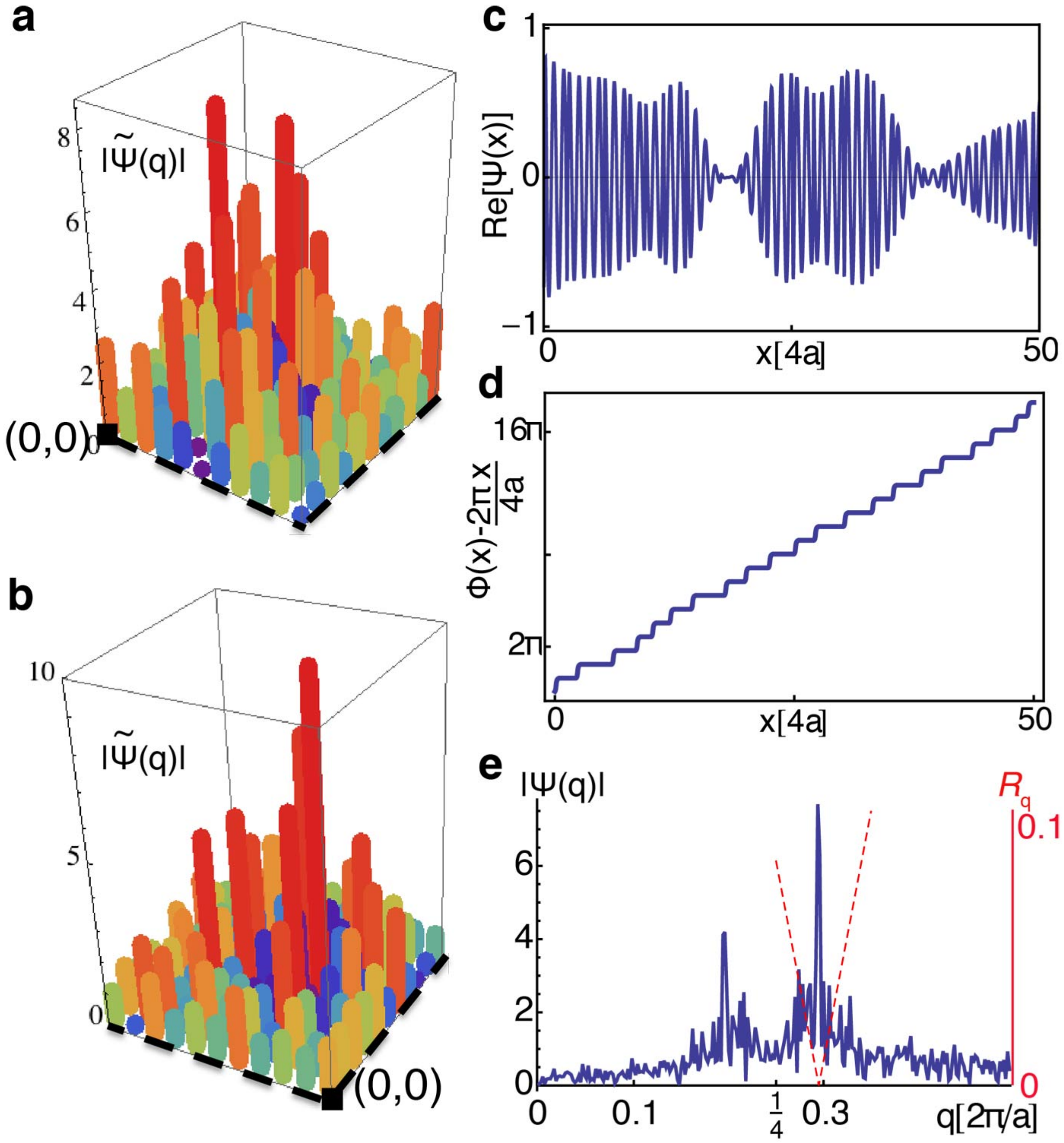
a
|Ψ̃(q)|
(0,0)
b
|Ψ̃(q)|
(0,0)
c
Re[Ψ(x)]
x[4a]
d
Φ(x)-2πx/4a
16π
2π
x[4a]
e
|Ψ(q)|
$R_q$
q[2π/a]
1/4
0.3

**Extended Data Fig. 10 Weakness of Fourier Transform analysis of EQM. a, b,** The DFF Fourier amplitude, $|\widetilde{\Psi}(\boldsymbol{q})|$, with wavevector $\boldsymbol{q}$ restricted to a square area with corner at the Fourier space origin (black square) and center at $\boldsymbol{Q}_X = \frac{1}{4}\boldsymbol{G}_X$ (in **a**) or $\boldsymbol{Q}_Y = \frac{1}{4}\boldsymbol{G}_Y$ (in **b**), where $\boldsymbol{G}_X$ and $\boldsymbol{G}_Y$ are the Bragg peaks. Data from $Bi_2Sr_2CaCu_2O_8$ sample at doping level *p=0.10* ($T_c$(K)=65K). **c,** Modulation is the real part of complex wave $\psi(x) = A(x)e^{i(Q_0\, x+\varphi(x))}$ having commensurate domains with local wavevector $Q_0 = \frac{1}{4}\times\frac{2\pi}{a}$ (period $4a$). The amplitude $A(x) \geq 0$ varies smoothly around value 1. Phase slips are incorporated in $\varphi(x)$ (see **d**). The average wavevector is $\bar{Q} = 0.3\times\frac{2\pi}{a}$. **d,** The local phase $\varphi(x)$ of $\psi(x)$ in **c**, constructed as a discommensuration (DC) array in the phase argument $\Phi(x) = Q_0\, x + \varphi(x)$. Phase slips of all DC's are set to $+\pi$. The distances between neighboring DC's vary randomly around average distance set by value of incommensurability $\delta = \bar{Q} - Q_0 = 0.05\times\frac{2\pi}{a}$ . **e,** Fourier amplitudes $|\tilde{\psi}(q)|$ of the modulation $\psi(x)$ in **c** (blue line) show narrow peak at $\bar{Q} = 0.3\times\frac{2\pi}{a}$. The demodulation residue $|R_q|$ (red dashed line) has the minimum exactly at the average $\bar{Q}$.